\documentclass[]{ar2e}
\usepackage{cite}
\newcommand{\etal}{{\it et al.}}

\newcommand{\eqref}[1]{(\ref{#1})}
\newcommand{\beq}{\begin{equation}}
\newcommand{\eeq}{\end{equation}}
\newcommand{\beqa}{\begin{eqnarray}}
\newcommand{\eeqa}{\end{eqnarray}}

\def\lsim{\raise0.3ex\hbox{$<$}\kern-0.75em{\lower0.65ex\hbox{$\sim$}}}
\def\gsim{\raise0.3ex\hbox{$>$}\kern-0.75em{\lower0.65ex\hbox{$\sim$}}}

\begin{document}

\input epsf.tex    
\input psfig.sty

\jname{..}
\jyear{2010}
\jvol{}
\ARinfo{1056-8700/97/0610-00}
\title{Supernova cosmology: legacy and future}

\markboth{Goobar \& Leibundgut}{SN cosmology: legacy and future}

\author{Ariel Goobar\affiliation{The Oskar Klein Center, Physics Department, Stockholm University,
    AlbaNova University Center, SE 106 192 Stockholm, Sweden, 
ariel@fysik.su.se} Bruno Leibundgut
\affiliation{European Southern Observatory, 
Karl-Schwarzschild-Strasse 2, 85748 Garching, Germany\\ and\\
Excellence Cluster Universe, Technische Universit\"at M\"unchen, Boltzmannstrasse 2, 85748 Garching, Germany, bleibund@eso.org}
}

\begin{abstract}
The discovery of {\em dark energy} by the first generation of
high-redshift supernova surveys has generated enormous interest beyond
cosmology and has dramatic implications for fundamental physics.
Distance measurements
using supernova explosions are the most direct probes of the expansion
history of the Universe, making them extremely useful tools to study
the cosmic fabric and the properties of gravity at the largest
scales. The past decade has seen the confirmation of the original
results. 
Type Ia supernovae are among the
leading techniques to obtain high-precision measurements 
of the dark energy equation of state parameter, and in the near 
future, its time dependence. The success of these efforts depends on
our ability to understand a large number of effects, mostly of
astrophysical nature, influencing the observed flux at Earth. 
The frontier now lies in understanding if the observed phenomenon is due
to vacuum energy, albeit its unnatural density, or some exotic new physics. 
Future surveys will address the systematic effects with improved
calibration procedures and provide thousands of supernovae for
detailed studies. 

\end{abstract}

\maketitle
\section{Introduction}
\label{intro}
Particle physics and cosmology are tightly linked and ideally
complement each other.  The physics of the early universe is beyond
the reach of even the most powerful accelerators and needs to be
deduced from the embers left by the Big Bang. The richness of the
cosmic microwave background radiation as a source of information of the early
universe has been described many times and has to be counted amongst
the most successful scientific endeavors of the past two decades
\cite{1999ARNPS..49...77K, 2002ARA&A..40..171H,
2007ARNPS..57..245S}. Observational cosmology can now address particle
physics questions, e.g. the number of neutrino species and the sum of
their masses, beyond the Standard Model particles as dark matter
candidates, and even braneworld scenarios invoking extra spatial
dimensions.  Theoretical high-energy physics has started to address
questions concerning the uniqueness of the observable universe and
String Theory inspired models to explain the current
state of the cosmos \cite{2005PhRvL..95n1301C,2006IJMPD..15.1753C}. 
Supernovae, the violent destruction of an
entire star, offer means to obtain accurate information about the expanding
universe, and are therefore a precious instrument in the cosmology tool box.
These luminous explosive events are important for modern cosmology
 since they can be detected even in very remote galaxies. Furthermore,
 some subtypes, Type Ia supernovae (SNe Ia) in particular, can be
 calibrated fairly reliably to provide accurate distances that can be
 used to map the expansion history of the universe.

The current expansion rate, the Hubble constant H$_0$, was determined
with supernova distances for several decades
\cite{1968AJ.....73.1021K, 1992ARA&A..30..359B, 1998ARA&A..36...17B}.
Two different methods have been employed using supernovae. Assuming a
unique luminosity for SNe Ia, often referred to as the ``standard
candle'' method, one can directly use the observed brightness to infer
the distance, an appealing technique due to its simplicity. Various
(astrophysical) effects influence the distance measurement, but
through suitable calibration and corrections this remains very
successful.  The other method is to measure the physical
expansion of a Type~II supernova through the radial velocity and its
brightness increase reflecting the growth in surface area. This method
has been applied to supernovae stemming from massive stars and with
extended envelopes.  In the local universe, this measurement is still
possible, but due to the limited luminosity of the Type~II 
supernovae becomes very difficult with current observing facilities at
larger distances
\cite{2006ApJ...645..841N}.

{\em Dark Energy}, detected using SNe~Ia through the accelerated
expansion of the universe, is amongst the most notorious recent
additions to our understanding of the cosmic composition
(e.g. \cite{2001ARA&A..39...67L, 2008GReGr..40..221L,
2008ARA&A..46..385F, 2009ARNPS..59..397C, 2010arXiv1011.0441H}). After
the first indications based on only small samples of supernovae
\cite{1998AJ....116.1009R, 1999ApJ...517..565P} the past decade has
seen a significant amount of telescope time invested in further
refining our understanding of SNe~Ia and their ability as distance
indicators. The cosmological results have been confirmed and the
uncertainties have decreased dramatically due to better
controlled observational techniques, improved calibrations, larger
samples and an extended redshift range. The current studies are
limited by the systematic uncertainties related to the supernova
explosions and the light propagation through the universe rather than
the size of the statistical sample.  Overall, there are now several
hundred Type Ia supernovae available for cosmic distance measurements
and there are prospects to increase these numbers at least ten fold
within the next decade. The field has moved from the discovery of dark
energy to its first characterization through the measurement of the
equation of state parameter, $w=p/\rho c^2$, i.e., the ratio between
pressure and energy density when treated as a fluid. Future
projects will address potential time-dependence of this parameter,
$w(t)$. Note that from now on, we will use natural units, i.e., $c=1$,
unless stated otherwise.

Although there is no observation today that is incompatible
with Einstein's cosmological constant, $\Lambda$, being responsible
for the accelerated expansion of the universe, only few high-energy 
physics theorists are ready to quit searching for alternative models.

A cosmological  constant ($w = -1$) can be naturally associated
with the vacuum zero-point energy predicted by quantum field theory.  
However, the corresponding density, $\rho_{vac} \sim (10^{-3} {\rm eV})^4$,
does not match any particle physics scale, and is more than 120 order
of magnitudes lower than the ``natural'' scale set by the Planck mass, 
$M_{P}^4 \sim (10^{28} {\rm eV})^4$
or
about 60 orders of magnitude below the supersymmetry breaking scale, 
$M^4_{SUSY} \sim (10^{12} {\rm eV})^4$.
Furthermore, it may be regarded as a great 
coincidence that  $\rho_{vac}$ and $\rho_M$, are so close 
in spite of $\rho_M \propto a^{-3}$, i.e.,
the density of non-relativistic matter gets diluted proportionally
to the growing volume of the expanding universe, while
the vacuum energy density remains constant. 

Some highly regarded physicists have invoked the Anthropic principle
to elude these difficulties arguing that we are witnessing the result
of a selection process: galaxies, stars, planets and intelligent beings would
only have a chance to come to being in ``un-natural'' universes like
ours \cite{1987PhRvL..59.2607W,1995PhRvL..74..846V}. Another common 
view on this problem is
that a yet unknown mechanism sets the vacuum energy to zero and the
present day acceleration is instead caused by new physics, e.g., a
light scalar field or an effective change of gravity at the largest
scales in a higher dimension universe. While the Anthropic hypothesis
cannot be further tested experimentally (with a possible exception
described in 
\cite{2006JCAP...05..009L}), we will not address it further here.
Alternatives to $\Lambda$, however, would imply that the
dark energy density would change in
time (and space), something that can be tested with SNIa, especially
when combined with other probes. In particular, measurements of the
expansion history and growth of structure in a universe with dynamically
evolving dark energy would lead to $w(t) \neq -1$, i.e., distinguishable  
from $\Lambda$.

A new generation of multi-probe projects are being planned, and their
relative merits have been assessed by the ESA--ESO Working Group on {\em Fundamental Cosmology} \cite{2006ewg3.rept.....P} and the {\em Dark Energy Task Force} report 
\cite{2006astro.ph..9591A}. These forthcoming measurements
are expected to provide the first accurate measurements of the 
time dependence of $w$.


The use of supernovae for cosmology has been reviewed many times.
The original cosmological application of supernovae was for the Hubble
constant \cite{1992ARA&A..30..359B, 1998ARA&A..36...17B,
2010ARA&A..48..673F}. The connection of supernovae with Dark Energy
has been described in many reviews as well \cite{2000PASP..112.1284R,
2001ARA&A..39...67L, 2003PhT....56d..53P, 2003LNP...598..195P,
2008ARA&A..46..385F}. Using supernovae to map the cosmic distance scale
requires a sufficient understanding of the explosions and the
astrophysical effects, which affect light propagation. These are
very important considerations for the derivation of accurate distances.

\subsection{Supernova classification}
\label{sec:class}

Supernovae have been recognised as an own class of astrophysical
objects by Baade and Zwicky \cite{1934PhRv...46...76B} following
earlier work by Lundmark \cite{1925MNRAS..85..865L}, who noted that
there is a class of eruptive variable objects, which are about 10
magnitudes (i.e. 10000 times) more luminous than regular (and more
frequent) novae. An early classfication scheme for supernovae was
introduced by Minkowski \cite{1941PASP...53..224M,
1964ARA&A...2..247M}.

Supernova classification is based on their spectral appearance near
maximum light 
\cite{1997ARA&A..35..309F}. The Type~II
supernovae show hydrogen in their maximum-light spectrum, while Type I
supernovae lack hydrogen. Type Ib SNe
display prominent helium lines, while the Type Ic display neither
hydrogen nor helium. Type Ia supernovae spectra are dominated by
lines from higher-mass elements, like calcium, sulphur, silicon and
iron, but lack hydrogen and helium - the most abundant elements in the
universe. 


The past years have seen some interesting modifications to the
classification after several decades of relative few new additions
\cite{1964ARA&A...2..247M, 1965stst.conf..367Z}. Several examples of
objects, which do not fit any of the established classes have been
added. The Type Ia class is overall very uniform in its appearance, but
shows some spread in its luminosity and spectral evolution. Both
overluminous and underluminous SNe~Ia have been identified (e.g.
\cite{2010arXiv1011.0441H}). In particular, a range of expansion
velocities -- as derived from the absorption trough of several lines
(e.g.  \cite{2009PASP..121..238B, 2006AJ....131.1648B,
2005ApJ...623.1011B, 2011A&A...526A.119N}) has been established.
These peculiar objects typically show interesting, and sometimes
subtle, deviations in their spectral appearance, which may allow to
refine their use as distance indicators (see Sec.~\ref{sec:precis}). 
Thanks to the large number of SN surveys conducted over the
past decade, a small number of truly different objects has been
discovered. Among them are objects, which appear very close to SNe~Ia,
but displayed a strong hydrogen emission (SN~2002ic;
\cite{2003Natur.424..651H}), extremely luminous objects with rather
slow expansion velocities (SN 2003fg; \cite{2006Natur.443..308H},
SN~2006gz; \cite{2007ApJ...669L..17H}, SN 2009dc;
\cite{2011MNRAS.410..585S}), and extremely faint objects (SN~2000cx;
\cite{2001PASP..113.1178L}, SN~2002cx; \cite{2003PASP..115..453L},
2005hk; \cite{2007PASP..119..360P}, 2008ha;
\cite{2009AJ....138..376F}, but see \cite{2009Natur.459..674V} for a
differing view on the classification of this object). In many cases,
the peculiar nature of these objects only became clear through a
detailed analysis of the pre-maximum color and spectral evolution or
the late phase evolution. It should be noted that the majority (70\%
in a volume-limited sample and 77\% in a magnitude-limited sample) of
SNe~Ia are quite homogeneous and only a few objects are true outliers
\cite{2010arXiv1006.4612L}.



The simple classification scheme of maximum light spectra has to be
expanded to include various additional parameters (color, light curve
shapes, etc.) for a complete description of the supernovae. The coming
years, with many large supernova surveys planned (Sec.~\ref{sec:precis}),
will provide the detailed data to define additional classification
criteria. 


Several major samples of supernovae have been assembled in the past
two decades. Leibundgut \cite{2000A&ARv..10..179L} provided an
overview of the situation a decade ago. In the meantime, the Center
for Astrophysics supernova program has produced an impressive sample of
around 160 nearby SNe~Ia \cite{2009ApJ...700..331H}, the Lick
Observatory Supernova Search has assembled several hundred SNe~Ia and
has published a first set recently \cite{2010ApJS..190..418G}, the
Carnegie Supernova Project has also collected a fairly large number of
nearby SNe~Ia with both optical and near-infrared data
\cite{2010AJ....139..120F}. The SNfactory has implemented a bold new technique 
for the study of supernovae: rather than carrying out a series of
photometry measurements in individual broadband filters, SNfactory
uses an integral field spectrograph to create synthesized photometry
corresponding to any possible optical filter set. The Palomar
Transient Factory\cite{2009PASP..121.1334R} and the PanSTARRS
supernova searches have started and have already produced a number of
interesting objects.  The distant searches will be described in
Sec.~\ref{sec:recent}. The nearby supernovae are critical for the
cosmology as they provide the comparison sample and also the anchoring
point of the distance scale.

\subsection{Supernova physics}
\label{sec:physics}


The display of a supernova is the result of the explosion mechanism
and the structure of the exploding star.
Collapsing about one solar mass from the radius of the Sun to a
neutron star or a black hole releases about 10$^{46}$~J, mostly in
neutrinos. About $10^{-2}$ of this energy goes into the acceleration of the
stellar material and $10^{-4}$ into radiation (10$^{42}$~J). 
The energy gain of burning about one solar mass to iron-group elements
due to the higher nuclear binding energy also is around 10$^{42}$~J.
Hence the two mechanisms produce very similar luminosities.

Type II and Type Ib/c SNe are associated with core-collapse supernovae
of massive stars \cite{2005NatPh...1..147W}, while the Type Ia
supernovae most likely are due to the thermonuclear explosion of a
compact white dwarf star \cite{2000ARA&A..38..191H}. 

\subsubsection{Thermonuclear explosions}
\label{sec:thermo}


The similarity between SNIa has been linked to a threshold phenomenon
involving at least one white dwarf accreting mass from the surrounding environment. These compact remnants of low mass
stars emit only thermal radiation from their slow gravitational contraction. 

\begin{figure}[h]
\epsfxsize=\textwidth
\centerline{
\epsfbox[1 1 500 350]{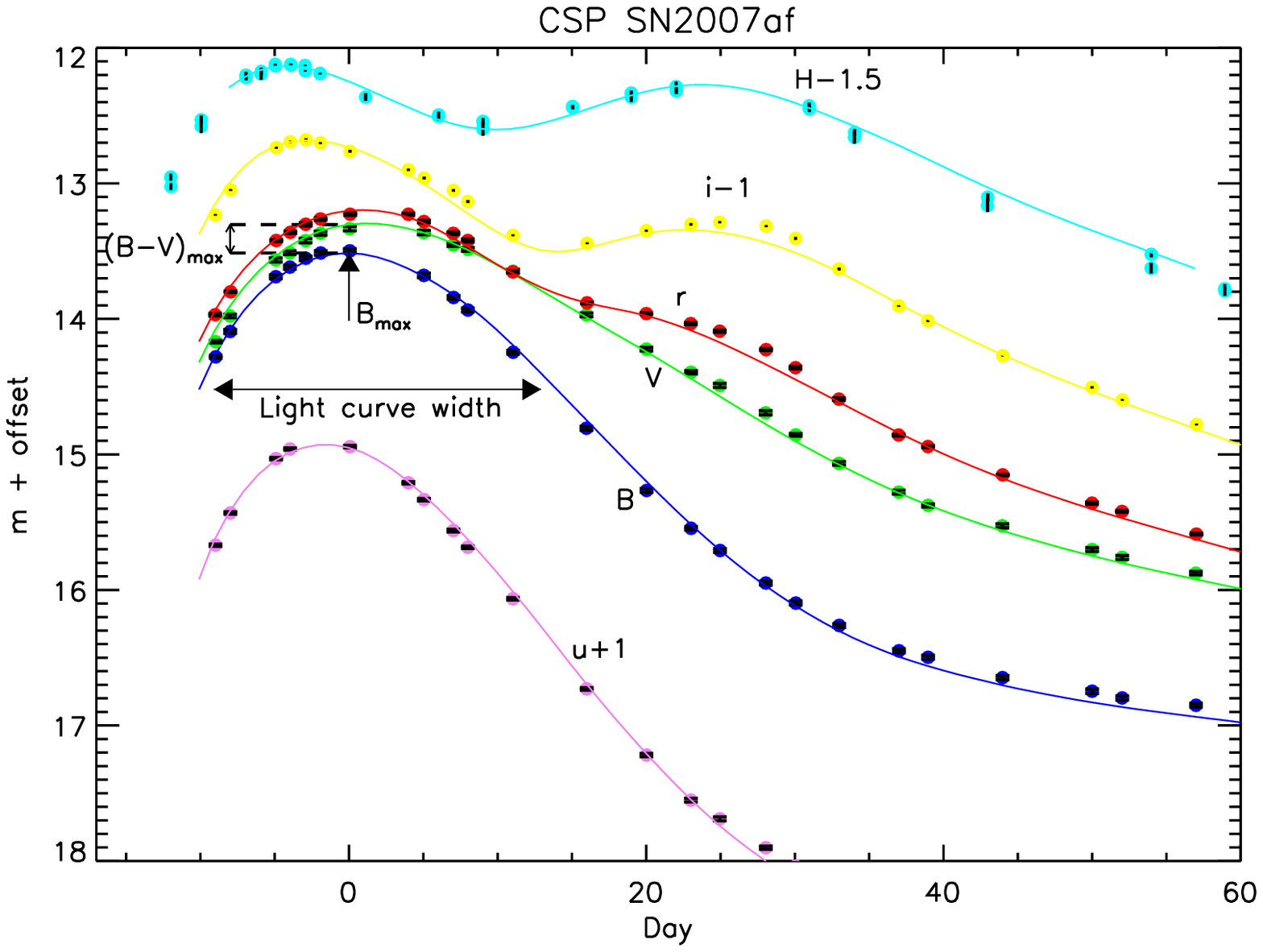}
}
\caption{
Optical and near-infrared light curves of the nearby SN~2007af
obtained by the Carnegie Supernova Project (Stritzinger \etal\ in
prep.). The light curves trace the brightness evolution in six
different wavelength passbands ranging from $\sim$3500\AA~(u) to
1.6~$\mu$m (H). For clarity the individual filter light curves have
been offset by the amounts indicated. Light curve fits
\cite{2011AJ....141...19B} are shown and the key fit parameters (time
and magnitude at maximum, the width of the light curve and a typical
color "B-V") are marked. The development of the second peak
with increasing wavelengths is very prominent. Note that the light curve
widths are used for comparisons within individual filters.
}
\label{fig:SN2007af}
\end{figure}

Thermonuclear explosions are due to the explosive burning of 
carbon and oxygen. The energy released from the synthesis of iron-group
elements is mostly used to overcome the binding energy to disintegrate
the star. The fairly quick brightness evolution (light curve) of Type
Ia supernovae (see Fig.~\ref{fig:SN2007af}), a rise time to maximum light of about 17 to 20 days and
a rapid decline after peak indicates a small progenitor star. 

The temperatures and densities for carbon burning are reached shortly
before the Chandrasekhar-mass, the highest mass a non-rotating white
dwarf can obtain before gravity overcomes the electron degeneracy and
collapses the white dwarf. Due to the electron degeneracy the
mass-radius relation is inverted and the more massive white dwarfs
actually have smaller radii. Once the carbon burning is triggered, it
starts a simmering phase within the star, which can last several hundred
years. During this phase the burning creates a large instability, which
at some point turns into a runaway (e.g. \cite{2000ARA&A..38..191H}). At
this point the burning proceeds through the star very rapidly and
disrupts it completely within seconds (e.g. \cite{2007ApJ...668.1132R}).
There are two ways the burning front can move through the star: at
subsonic speeds, called a deflagration, or at supersonic velocity,
designated detonation. Pure deflagrations typically have the problem
that they do not achieve enough energy for the bright display of Type Ia
SNe, but they have been suggested for several of the less luminous
SNe~Ia. Detonations have been disfavored for many years as they burn
most of the star to iron-group elements and do not leave a significant
amount of intermediate-mass elements, like silicon, calcium and sulphur
observed in the spectra. Models with a transition from
deflagration to detonation have been favored.  In this case, the
subsonic burning front accelerates to reach supersonic speed in the
outer layers and hence provides a mix of iron-group and
intermediate-mass elements. The exact reason for the transition,
however, has not been completely identified. 

Mass accretion is critical in this scenario. The white dwarf can only
grow in mass, if it receives matter from a companion star. This
immediately requires a close binary stellar system. Also, the white dwarf
needs to grow to close to the Chandrasekhar mass ($\sim 1.4 M_{\odot}$), which is not easy
given that typical white dwarfs are normally below a solar mass.
Possible accretion scenarios, both with significant problems at the
moment, are either the merging of two white dwarfs (e.g.
\cite{2010Natur.463...61P} for at least sub-luminous SNe~Ia) or mass
loss from the nearby companion star. There are observational arguments
for both these channels. A further option is to trigger the explosion
well below the Chandrasekhar mass by explosive burning of surface
material.  In such a case, helium detonates near the surface, possibly
where the accretion reaches the white dwarf, and then triggers the central
carbon explosion by pressure waves focussing near the center. While this
possibility has for a long time been regarded as producing the wrong
nucleosynthesis, it has recently been revived in full 3-dimensional simulations
\cite{2007A&A...476.1133F, 2010A&A...514A..53F}.

The optical display of a Type Ia supernova comes from the radioactive
decay of newly synthesized material. Adiabatic cooling of the ejecta
means that there is no remnant energy for optical emission. Instead
the decay chain from $^{56}$Ni$\rightarrow ^{56}$Co$\rightarrow
^{56}$Fe (e.g. \cite{1998PASP..110..637D}) is responsible for powering
the light curve. The radiation transport in Type Ia supernovae is very
complicated as the original $\gamma-$rays from the radioactive decay
are down-scattered to lower energies. This happens mostly through
Compton scattering and then absorption and line emission in local
atoms (e.g. \cite{2009MNRAS.398.1809K}).  This non-thermal process is
extremely difficult to model and depends on complete atomic data for
many ionized atoms. The emission at different wavelengths is modulated
by these radiation transport effects. A striking example is the
secondary peak, which is observed in the light curves of red and
infrared filters (at $\lambda \geq 700$nm, see Fig.~\ref{fig:SN2007af}). This has been 
tentatively explained as
due to the ionisation state of iron \cite{2006ApJ...649..939K}, where
the opacities change dramatically for doubly- to singly-ionised iron.  

The original hopes of a single explosion mechanism for all SNe~Ia is
probably no longer tenable. The observed diversity indicates that
several of the evolutionary scenarii and also explosion mechanisms are
at work. It has also emerged that the explosions are unlikely to be
spherical and hence viewing angle effects need to be taken into account
as well \cite{2009Natur.460..869K}. In particular, they can introduce a
natural scatter into the observables. 

Critical observational input to the models would be the total mass of
the explosion, the newly synthesized $^{56}$Ni mass and the explosion
energy (e.g. \cite{2006A&A...450..241S}). This information is difficult
to come by and is being assembled only slowly.

\subsection{Supernova rates}

Supernovae are rare objects. The last millennium has only six confirmed
supernovae in our own Milky Way and SN~1987A in the Large Magellanic
Cloud. The local rate of Type Ia supernovae is about
$3\cdot10^{-5}$~SNe~Ia Mpc$^{-3}$yr$^{-1}$
\cite{2010arXiv1006.4613L, 2010ApJ...713.1026D}.  
The past decade has seen the industrialization of sky monitoring and
supernovae are now found regularly. The challenge is different for
nearby and distant supernovae.  Due to the limited volume it is rather
difficult for find nearby supernovae as large sky areas need to be
monitored. Nevertheless, the rate of supernova discoveries in the
local volume (out to about 100~Mpc) has doubled in the past
decade. For higher redshifts, the increase has been much larger due to
the larger mirror/longer exposures and wide-area capabilities, since
distant supernovae are found by deep pencil-beam searches. Due to the
great depth, a large volume is searched and many supernovae are found.

\begin{figure}[h]
\epsfxsize=\textwidth
\centerline{
\epsfbox[50 350 400 700]{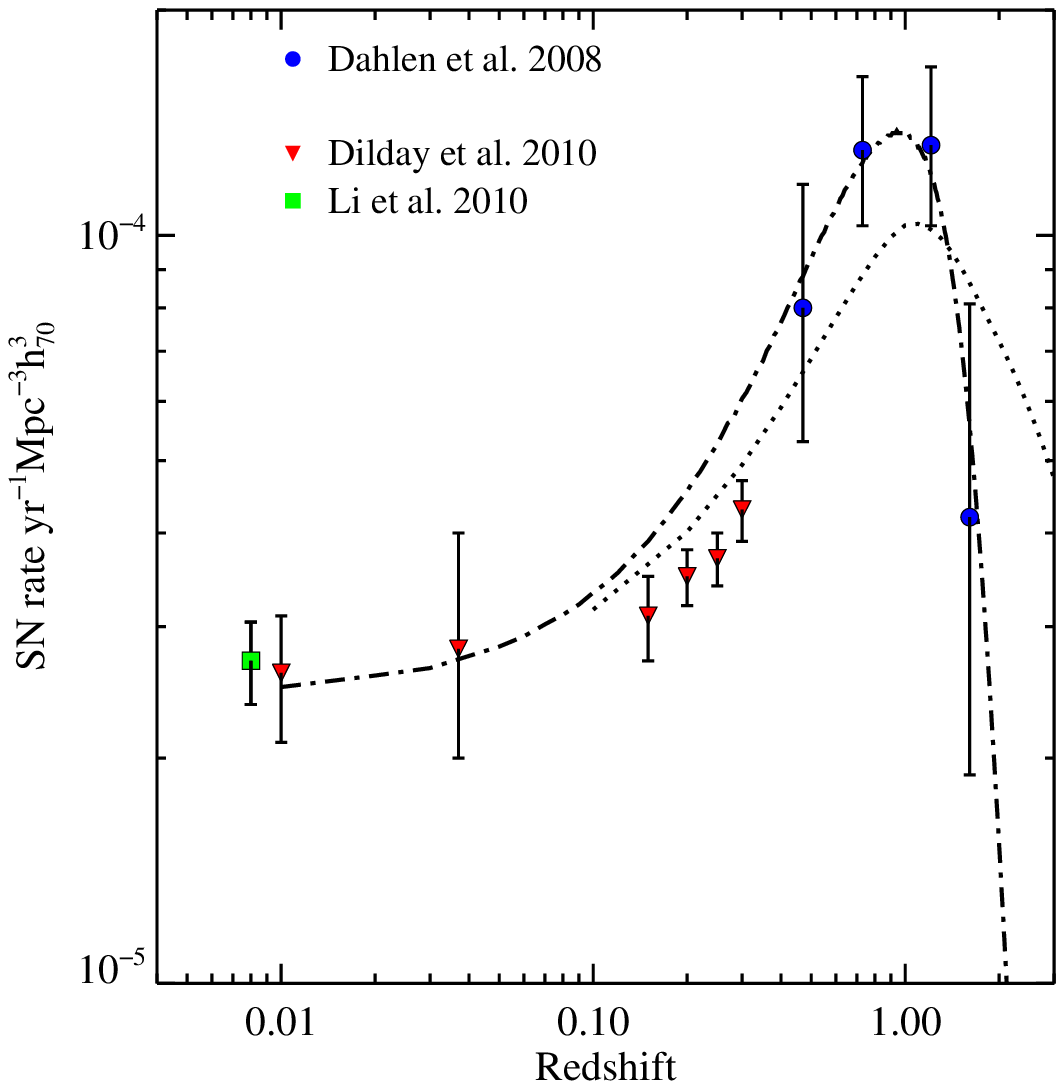}
}
\caption{
The evolution of the Type Ia supernova rate per unit volume as a
function of redshift (data from \cite{2008ApJ...681..462D}). The
error bars reflect the statistical uncertainties per redshift bin. The
highest redshift bin contains 3 objects. The
increased rate with redshift follows the increased star formation
rate observed in the universe. The two models
\cite{2004ApJ...613..189D,2007MNRAS.377.1229M} show the effect of
different progenitor life times. A long progenitor evolution towards a
supernova (dash-dotted line) means that no SNe~Ia are observed in the
early universe. In the case of a shorter delay time (dotted line) the
rate remains high at high redshifts (and large look-back times). 
}
\label{fig:rates}
\end{figure}

The SN~Ia rate is connected to the star formation of the local galaxy
and can also provide information on the gestation time before a
supernova explosion occurs. The progenitor systems of SNe~Ia have not
been uniquely identified and hence the rate as a function of redshift
(or look-back time) provides important clues. For systems, which take
a long time to grow the white dwarf towards the Chandrasekhar mass
(several 10$^9$ years) there has to be a cutoff when the universe is
not old enough to produce a Type Ia supernova. Figure~\ref{fig:rates}
displays the current state of affairs. There is clearly an increase by
a factor of 3 in the SN~Ia rate back to a redshift of 1. If the
drop for z$>$1 is real then we may see the onset of the first
supernovae at that time. 

Searches at very high redshifts are currently under way (as part of
the CLASH and CANDELS HST multi-cycle projects). They will show
whether indeed there are no Type Ia supernovae at $z=2$ (about 3 billion
years after the Big Bang for the canonical cosmological parameters.

\section{Standard cosmology}
\label{sec:cosmo}

Modern cosmology is based on Einstein's field 
equations of General Relativity (GR):
\beq
R_{\mu\nu}-{1\over 2}g_{\mu\nu}R={8\pi G}T_{\mu\nu} + \Lambda g_{\mu\nu}, 
\label{eq:einstfe}
\eeq which connects spatial curvature (left hand side) and energy
content (right hand side). Equation \ref{eq:einstfe} includes the
extra term $\Lambda$, also called the {\em cosmological constant},
that Einstein introduced in 1917 to account for the possibility that,
 on the largest scales of the universe, the attractive nature
of gravity could be neutralized, thus potentially allowing for a
static solution. In the context of quantum field theory, a $\Lambda$
term arises naturally as a vacuum energy density, 
$\rho_{vac}= {\Lambda \over 8 \pi G}$, i.e. the zero-point energy 
associated with the production and annihilation of all virtual
particles. 

\subsection{Friedmann-Lema\^itre models}
\label{sec:Friedmann}

To compute cosmological observables from GR, 
Eq.~(\ref{eq:einstfe}) is solved using two simplifying assumptions about
the universe: {\em homogeneity} and {\em isotropy} over very large
scales. Expressed in the Robertson-Walker line element, distances
in space-time become ($c=1$):
\beq
ds^2 = dt^2 - a(t)^2 \left( {{dr^2}\over{1- kr^2}} + r^2 d\theta^2 + 
       r^2\sin^2\theta  d\phi^2 \right),
\label{eq:RW}
\eeq
where $a(t)$ is scale factor of the expanding universe and $k$ indicates
the curvature: $(0,1,-1)$ for a flat, closed or open universe. 

The solutions to the temporal and spatial components of the  
Einstein's field equations in the Robertson-Walker
metric lead to two fundamental equations for the dynamics of
the universe first found by Friedmann and Lema\^itre 
for the first and second derivative
of the scale factor, $a$:

\beq
\left( {\dot a \over a} \right)^2 = {{8 \pi G} \over {3}} \rho - {k \over a^2} 
\label{eq:friedmann}
\eeq
and \\
\beq
{\ddot a \over a} = -{{4 \pi G} \over {3}} (\rho + 3 p),
\label{eq:acceleration}
\eeq

where $\rho$ and $p$ are the energy density and pressure of the
cosmic fluids, and are typically related through the equation of 
state parameter as:
\beq
p = w \cdot \rho 
\label{eq:eqofstate}
\eeq

Since the discovery of the expansion of the universe in the late 1920's, 
the current expansion rate is called the Hubble constant, 
${\dot a \over a}=H_0$. The goal of observational cosmology since
then, is to measure the time dependence of the expansion rate, $H(t)$, thus indirectly
probing the contents of the universe.

\subsection{Components in the Friedmann-Lema\^itre cosmological models}
\label{sec:FLcomponents}

In the Standard Model of cosmology, the universe has been expanding and
cooling since almost 14 billion years ($\sim H_0^{-1}$). Thus, after the
initial ``Big Bang'', the energy content of the universe was dominated 
by relativistic particles ($w=1/3$), eventually surpassed by non-relativistic
matter ($w=0$) as the temperature fell well below the mass of standard
model particles.\footnote{A convenient relation between age of the universe
and the plasma temperature in the early universe is $T({\rm MeV}) = t^{-{1 \over 2}}({\rm s})$.}

A continuity equation can be derived from 
Eqs.~(\ref{eq:friedmann}-\ref{eq:eqofstate}):

\beq
\dot \rho + 3 H \rho (1 + w) = 0
\label{eq:continuity}
\eeq

The density of cosmic fluids in the expanding universe evolves according
to Eq.~(\ref{eq:continuity}) as:

\beq
\rho = \rho_0 e^{3 \int_{a_{0}}^a {{1 + w} \over {a'}} da'},
\label{eq:density_gen}
\eeq
where $a_0$ is the scale factor today.

A simple relation is thus found for fluids with {\em constant} $w$:
\beq
\rho = \rho_0 \left( {a_0 \over a} \right)^{3(1+w)}.
\label{eq:density}
\eeq

We thus find the familiar dilution for non-relativistic matter as the
universe expands, $\rho_M = \rho_M^0 \left(a_0 \over a\right)^3$, and
because of redshift, by another power of $a$ for relativistic matter
$\rho_R = \rho_R^0\left(a_0 \over a\right)^4$. Since $\Lambda$ is
constant, is is directly seen from Eq.~(\ref{eq:density}) that its
equation of state is $w_\Lambda=-1$. More generically, any dominant
fluid with $w<-1/3$ will cause the universe to accelerate, as seen
from Eq.~(\ref{eq:acceleration}).  Such fluids are called {\em dark
energy}, the ``simplest'' one being the vacuum energy density
associated with the cosmological constant, $\Lambda$. Other dark
energy candidates are discussed in e.g.~\cite{2006IJMPD..15.1753C,2009ARNPS..59..397C}.

It is customary to re-write the Friedmann equation (\ref{eq:friedmann}) 
in terms of the cosmological redshift, $1+z={a_{0} \over a}$, and the
density terms normalized by the critical density corresponding to a 
flat universe, $\Omega_{DE}={\rho_{DE} \over \rho_c} = 
             \rho_{DE} \cdot {8 \pi G \over 3H_0} $, and the curvature term is defined as $\Omega_K = -k(a_0H_0)^{-2}$. 

Since we lack the knowledge of the nature and properties of the equation of
state of dark energy, phenomenological parameterizations
of $w(z)$ are used for comparisons with data, e.g.
\beq
w(z) = w_0 + w_a \left(1 - {{a} \over {a_0}} \right) = w_0 + w_a {z \over 1+ z},
\label{eq:w0wa}
\eeq
where the free parameters $w_0$ and $w_a$ are fitted with observational
data.


\subsection{The luminosity distance and the Hubble diagram}
\label{sec:Hubble}
\begin{figure}[h]
\epsfxsize=\textwidth
\centerline{\epsfbox[1 170 640 660] {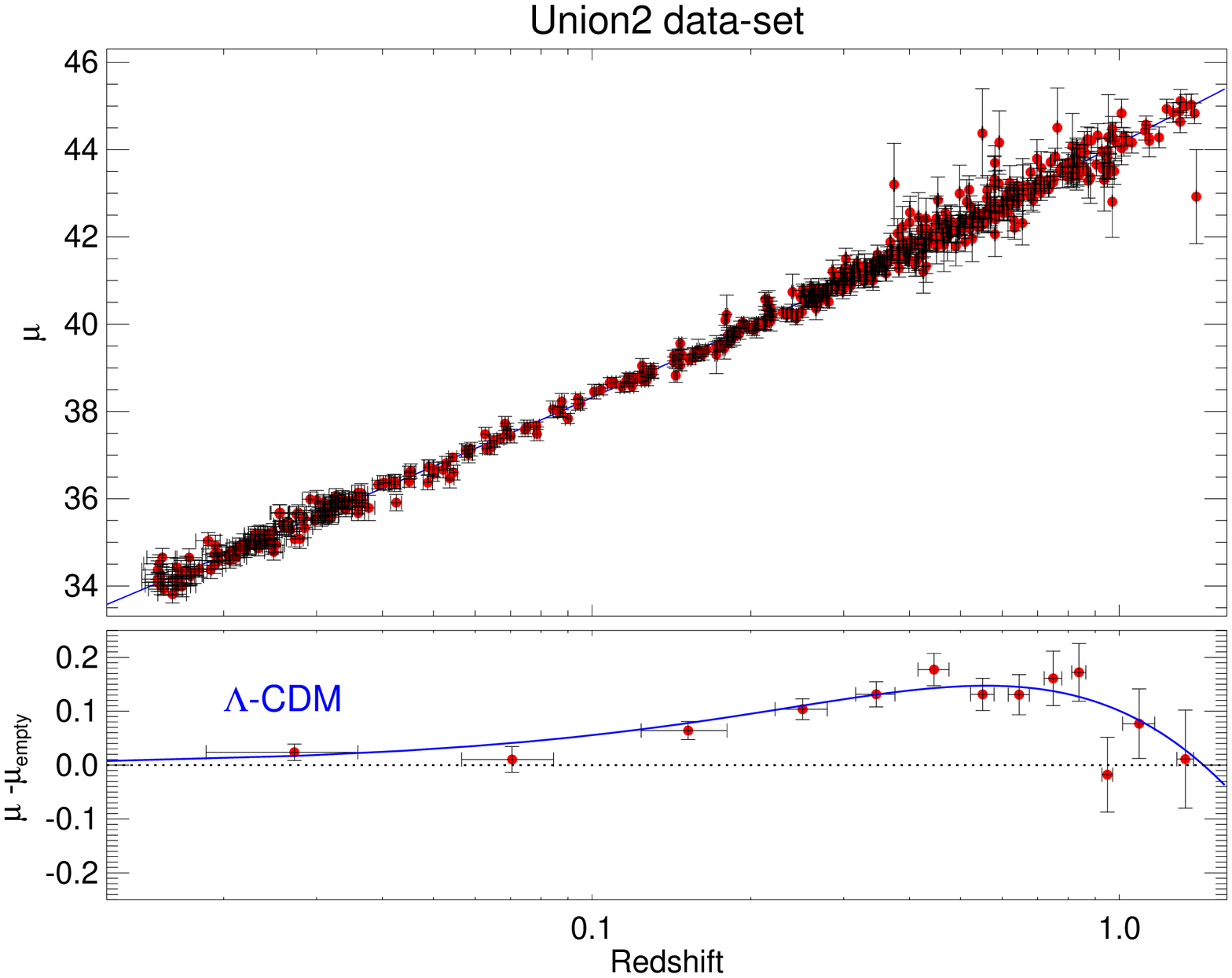}}
\caption{
The Hubble diagram of Type Ia supernovae correlating distance modulus
($\mu$) $vs.$ redshift. The Union2 compilation
\cite{2010ApJ...716..712A} represents the currently largest SN~Ia
sample. The linear expansion in the local universe can be traced out
to z$<$0.1. The Hubble constant sets the absolute level of the data,
while it is irrelevant for the determination of $\Omega_{\Lambda}$,
which is a relative measurement (bottom panel). The distance relative
to an empty universe model ($\mu_{empty}$; $\Omega_M =
\Omega_{\Lambda} = 0$) is shown in the lower panel. The data are
binned for clarity in this diagram. The blue curve shows the
expectation from the best fit $\Lambda$CDM model ($\Omega_M=0.3,
\Omega_{\Lambda}=0.7$).
}
\label{fig:Union2}
\end{figure}

There are several cosmological distances measures. Here we 
concentrate on the {\em Luminosity Distance}, $d_L$, which is defined
from the expected flux, ${\cal F}$, from an object with luminosity $L$
(in units of energy per unity time and unit area), ${\cal F}={L \over
4 \pi d_L^2}$.  Since the universe expands, any cosmological distance
estimate will depend on the growth of the scale factor and thus on the
Hubble parameter. For a full derivation, the reader should consult a
cosmology text book, e.g. \cite{1999cpa..book.....B}. Here we
concentrate on the final expression (neglecting the radiation density
term $\Omega_R \sim 10^{-5}$):

\begin{eqnarray}
 \label{eq:dl1}
    d_L&=&\left\{
    \begin{array}{ll}
      (1+z)\frac{1}{\sqrt{-\Omega_K}}\sin(\sqrt{-\Omega_K}\,I) , &
      \Omega_K<0\\
      (1+z)\,I , & \Omega_K=0\\
      (1+z)\frac{1}{\sqrt{\Omega_K}}\sinh(\sqrt{\Omega_K}\,I) , &
      \Omega_K>0\\

    \end{array}
    \right. \\
    \Omega_K&=&1-\Omega_M-\Omega_{DE},\\
    I&=&\int_0^z\,\frac{dz'}{H(z')} ,\\
    H(z)&=& H_0\sqrt{(1+z)^3\,\Omega_{M}+
    f(z)\,\Omega_{DE}+(1+z)^2\,\Omega_{K}}, \\
    f(z)&=&\exp\left[3\int_0^z\,dz'\,\frac{1+w(z')}{1+z'}\right].
 \label{eq:dl} 
\end{eqnarray}
It is customary to use logarithmic distances in astronomy, the {\em
distance modulus}, $\mu = 5\log_{10}d_L({\rm Mpc}) + 25 = m_f - M_f$,
where the last term corresponds to the difference between the measured
magnitude of the object, $m_f$ and its absolute magnitude through the
same filter $f$, $M_f$, i.e. corresponding to the flux of the object
if placed 10 pc away from the observer.  Cosmological model testing is
done by comparing measured distance moduli {\em vs.} redshift against model
predictions. This is usually called the Hubble diagram. The supernova
Hubble diagram, which will be discussed in more detail in the
following is displayed in Fig.~\ref{fig:Union2}. 

\section{Cosmology with Type Ia supernovae}
\label{sec:Ia_cosmo}

Since SNe~Ia display rather uniform light curves, they can be used as
local clocks. This test for cosmological time dilation had been
proposed soon after supernovae were recognized and was implemented with the first distant
supernovae \cite{1996ApJ...466L..21L, 2001ApJ...558..359G}. The
spectral evolution also tests for this stretching of time in
the observer frame \cite{2008ApJ...682..724B}. All tests so far have
confirmed the prediction of general relativity. The corrections for
time dilation are now routinely applied to the supernova data. 

\begin{figure}[h]
\epsfxsize=\textwidth
\centerline{\epsfbox[1 1 500 150]{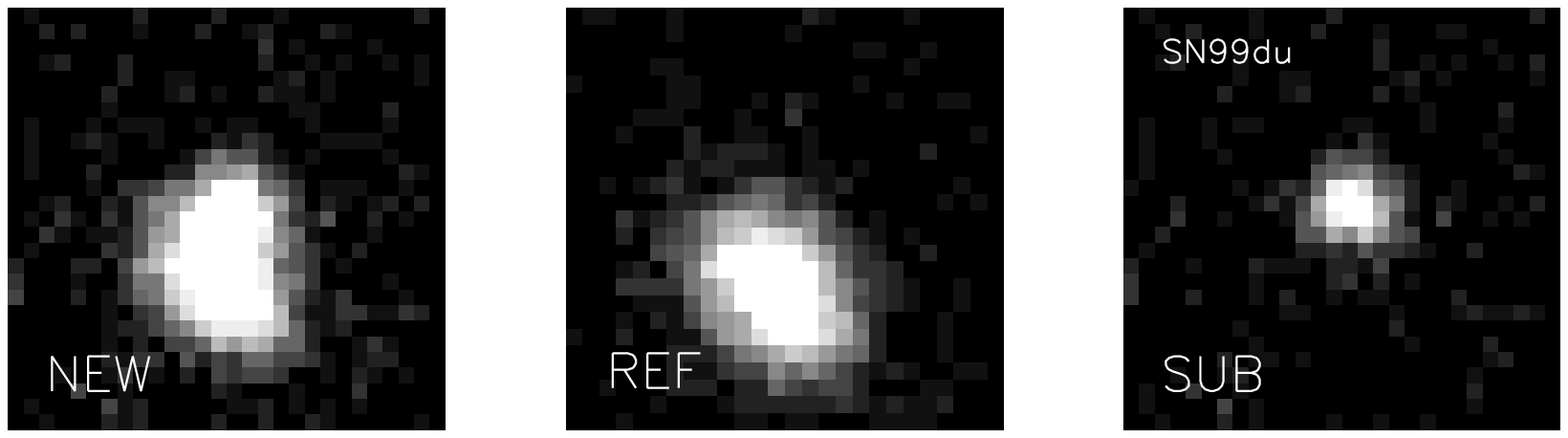} }
\caption{Detection principle for supernovae: images separated by between 
a few days and weeks are subtracted from each other (NEW - REF = SUB). 
In this case, $g$-band images used to detect SN~1999du ($z=0.260$) in 
\cite{2008A&A...486..375A}
are shown.}
\label{fig:subtraction}
\end{figure}

Most supernova searches operate by repeated imaging of patches of the
sky with many thousand galaxies. Supernovae are found through
pixel-by-pixel subtractions of CCD images obtained at different
epochs, as shown in Figure \ref{fig:subtraction}. 


\begin{figure}[h]
\epsfxsize=\textwidth
\centerline{\epsfbox[1 70 480 330]{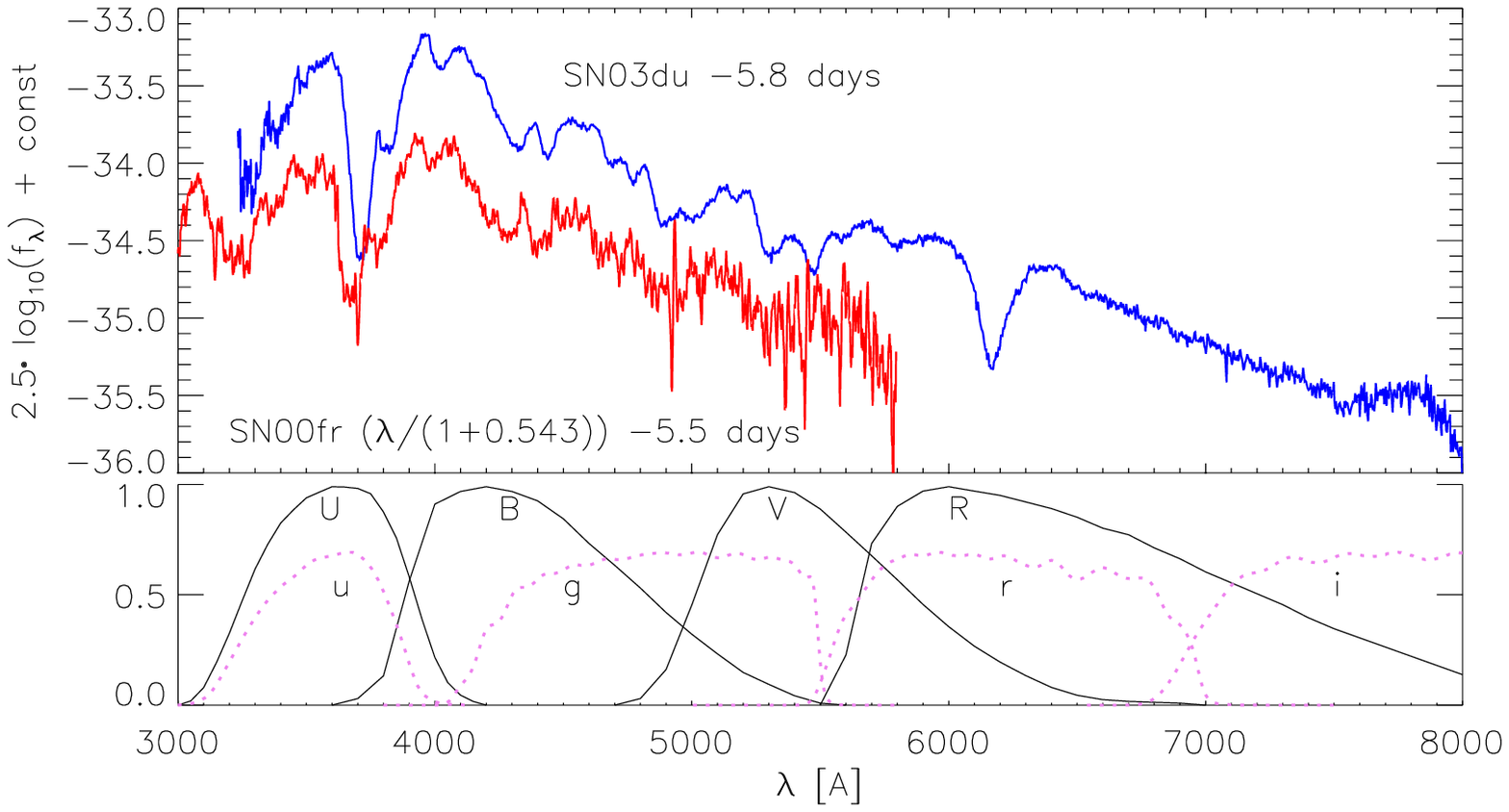}  }
\caption{
The observed spectrum of SN~2003du \cite{2007A&A...469..645S} compared
with the restframe spectrum of SN~2000fr (z=0.543;
\cite{2005A&A...430..843L}), both about 6 restframe days prior to
B-filter maximum. The two spectra display remarkable similarities in
the absorption and emission regions. Currently available data have not
provided any evidence of clearly different spectral appearance of
distant SNe~Ia compared to their local counterparts indicating that
there are no major changes in the explosion physics. The spectrum of the nearby
supernova extends to much longer wavelengths, which are redshifted
outside the observable window for the distant supernova, where the
restframe UV region (not fully shown) is covered much more
extensively. The bottom panel indicates the optical filter bands for
the Johnson-Cousins UBVR system as well as the SDSS ugri filter set. 
}
\label{fig:spectrum}
\end{figure}

\label{sec:lightcurve}
Once a supernova is detected, a few steps are necessary before the
brightness can be used for distances estimates. First and foremost,
the object is spectroscopically classified, as described in
Sec.~\ref{sec:class}.  The spectrum also provides the vital redshift
for the object. In many cases light from the host galaxy can be
analyzed to find a redshift, but for ``hostless'' supernovae 
(where the surface brightness of the host galaxy falls below
the detection limit) the redshift
can also be determined from the supernova spectrum itself. Figure
\ref{fig:spectrum} shows the identification spectra of two Type Ia
supernovae, SN2003du in the nearby universe, and SN2000fr at
$z=0.543$, i.e., more than 8 Gyrs apart.  The striking spectral
similarity is the key of the ``standard candle'' technique.  Since
distances are established based on the peak flux, light curves are
built through several filters, ideally including observations well
before and after maximum, as shown in Fig.~\ref{fig:SN2007af}. Besides time dilation making the light curve
width increasingly broader with redshift, SNIa optical light curves
exhibit a range in widths. The correlation between light curve and
supernova luminosity has been established over several years. More
luminous supernovae display typically a slightly slower evolution.

\begin{figure}[h]
\epsfxsize=\textwidth
\centerline{\epsfbox[1 1 700 700]{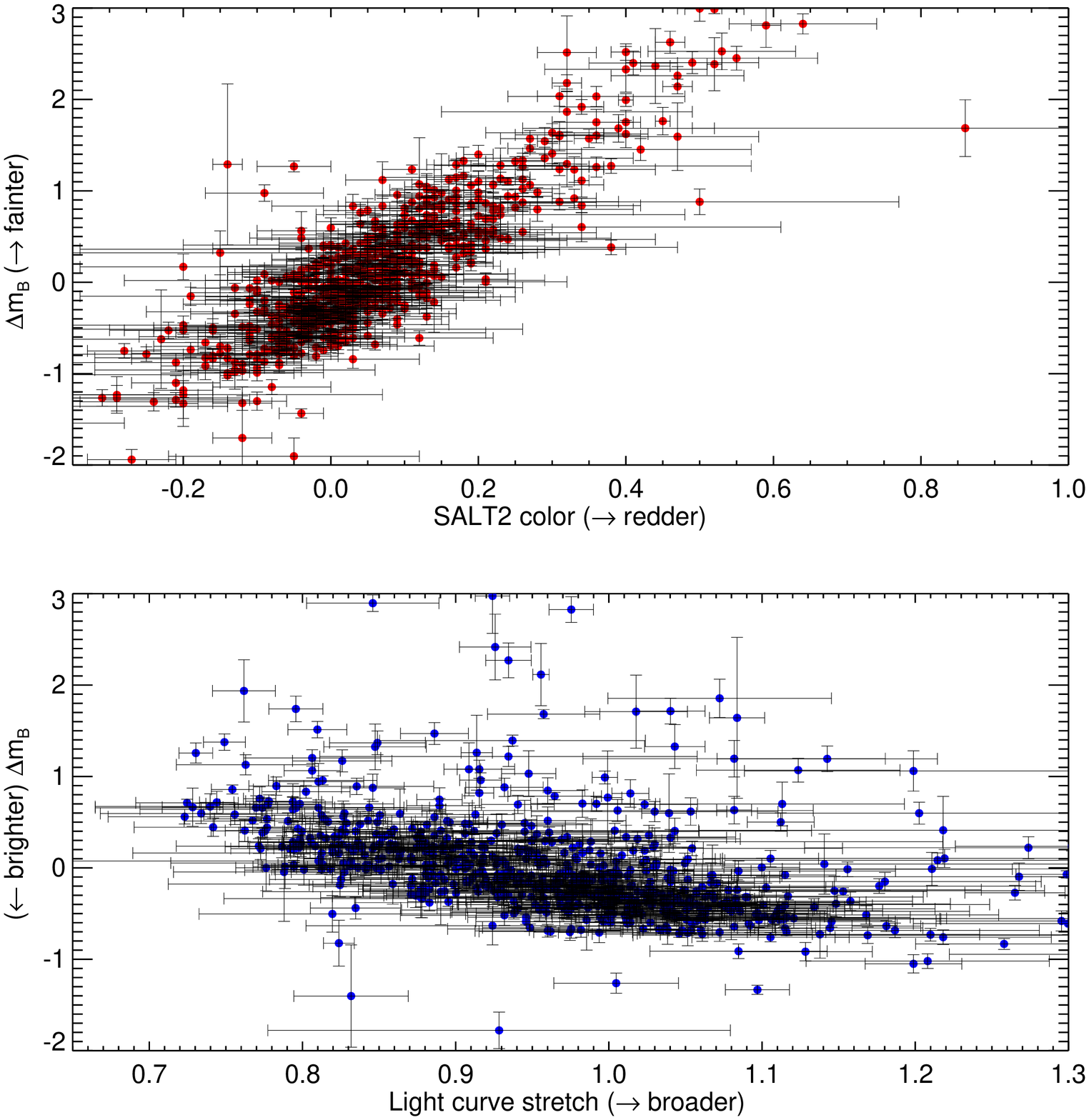}  }
\caption{
The color $vs.$ luminosity (upper panel) and the light curve shape $vs.$ 
luminosity (lower panel, after color corrections) correlations of Type Ia supernovae. Light
curve fits to the restframe B filter observations of 685 SNe~Ia
\cite{2010ApJ...716..712A} 
covering a redshift range from 0.025$<$z$<$1.4 are displayed. The clear
correlations are used to correct the distances and provide a
significant reduction in the scatter. It should be noted that the
color has already been corrected for Milky Way reddening.   
}
\label{fig:SNcalibration}
\end{figure}

 
Intervening dust can scatter and absorb supernova light and hence produce a
dimming, which needs to be corrected for the distance measurement.
Additional effects may also influence the observed colors of SNIa, as
discussed in Section ~\ref{sec:precis}. Several light curve ``fitters''
have been proposed to extract the key parameters from the data: the
day of maximum and the corresponding peak flux, the light curve shape,
and the reddening (color excess). The latter is often computed as the
relative magnitude attenuation between the restframe $B$-band and the
$V$-band. The empirical {\em color-brightness} and {\em lighcturve 
shape-brightness} relations used to ``standardize'' SNe~Ia
are shown in Fig.~\ref{fig:SNcalibration}.

Since the measurements are done with fixed filter pass-bands,
irrespectively of the supernova redshift, the measured fluxes
correspond to different parts of the restframe spectrum. 
The distance estimates rely on flux ratios of the same restframe
filters and hence, a $K$-correction is
needed to transfer the observed flux into the restframe for 
comparison with local objects.
Due to the significant spectral evolution, 
high S/N spectroscopic observations would be needed for every
flux point, a daunting task! However, given the uniformity
of SNe~Ia, spectral {\em templates} are used to compute
$K$-corrections. These templates are made of well sampled, low noise,
measurements of SNe~Ia, mostly at low redshift. Any potential
reddening is determined by comparison of
the multi-band data with synthetic colors calculated from
spectral templates after applying an extinction law, e.g., the
one found for stars in the Milky Way \cite{1989ApJ...345..245C}. The
photometric measurements from different telescopes also need to be
corrected for differences in the wavelength response of the telescope
reflectivity and instrument sensitivity. Modern analysis methods
try to correct for these systematic effects in the light curve
fitting. 

To summarize, observed SN peak magnitudes at redshift $z$ through a
filter $Y$ are compared with model predictions taking into account the
observed color and light curve shape ($c,s$).  The effect of
cosmological redshift on the spectrum, from the restframe wavelength
$X$ to the observations in $Y$ are handled through the cross-filter
$K_{XY}$-corrections \cite{1996PASP..108..190K}. These would typically
also include effects from reddening and light curve shape. 

\begin{equation}
m_Y(z,d,c) = M_X(s) + 5 \log_{10} d_L(z) + K_{XY}(z,s,c) + A^{MW}_{Y}, 
\label{eq:mxy}
\end{equation}

where the last term corresponds to dimming by Milky-Way dust. Note that
in Eq.~(\ref{eq:mxy}) it is implicitly assumed that corrections for 
extinction in the host galaxy are included in $K_{XY}$.

%
%

\subsection{The Hubble constant}
As discussed in Sec.~\ref{sec:Hubble}, the relation between the distance
modulus and cosmological redshift can be used to measure cosmological
parameters such as $H_0$, $\Omega_M$, $\Omega_{DE}$ and the dark energy
equation of state parameter, $w$. For $z \ll 1$, we find from
Eq.~(\ref{eq:dl}):
\begin{equation}
d_L = {cz \over H_0} \left(1 + {1 \over 2}(1-q_0)z + {\cal O}(z^2)\right),
\label{eq:dllowz}
\end{equation}
where we have explicitly reinserted the speed of light $c$ into the 
expression and introduced the deceleration parameter, 
$q_0$, which is related to the cosmological ``fluids'' as:
\begin{eqnarray}
q_0 & = & -\left({\ddot a \over a}\right){1 \over H_0^2} \\
    & = & \sum_i {\Omega_i \over 2} \left(1 + 3{p_i\over \rho_ic^2}\right).
\end{eqnarray}
The linear part of the Hubble diagram, $d_L(z\ll 1) \approx cz/H_0$,
has been used to measure $H_0$ since several decades using 
magnitude measurements to SNIa using optical data, e.g., through the
$B$-band filter, $m_B$:
\begin{equation}
\mu = m_B - M_B = 5 \log{cz} - 5\log{H_0} + 25 + K_{BB} + A^{MW}_B.   
\end{equation}
Thus, $\log{H_0}$ can be read off from
the intercept of the $\log{cz}-0.2m_B$ relation (cf.
Fig.~\ref{fig:Union2}).


While SNe Ia are very useful for measuring relative distances,
extracting $H_0$ requires an absolute calibration of the SNIa
brightness. Since theoretical models are not sufficiently accurate,
the SNIa absolute magnitude needs to be calibrated observationally
\cite{2010ARA&A..48..673F}. Cepheid stars are used to measure
distances to galaxies in the nearby universe. By now, Cepheid
distances to 15 SNIa host galaxies within about 20 Mpc have been
obtained with the {\em Hubble Space Telescope}
\cite{2001ApJ...553...47F}.  Riess \etal 
\cite{2009ApJ...699..539R} mitigated some of the shortcomings in
building up the {\em distance ladder} to obtain absolute distances to
SN~Ia hosts. Near-IR observations of Cepheids in six systems with
accurate SN~Ia data were used to improve the calibration and reducing
the uncertainty from extinction by dust. Furthermore, a better
anchoring system was found to calibrate the Cepheid distances, the
``maser'' galaxy NGC 4258, for which an accurate absolute distance can
be derived.  Riess \etal \cite{2009ApJ...699..539R} report $H_0= 74.2
\pm 3.6$ km s$^{-1}$Mpc$^{-1}$, where the 4.8 \% error includes
systematic uncertainties.

\subsection{The accelerated expansion}
\label{sec:DE}
The extension of the SNIa Hubble diagram to higher redshifts, needed
to probe the change in the cosmic expansion rate, requires a specific
observational strategy: large patches of the sky need to be scanned
regularly. 
This prompted the need for medium size telescopes, $2-4$ meters in
diameter, equipped with CCD cameras with large fields of view. 

By the mid-nineties, Goobar \& Perlmutter  \cite{1995ApJ...450...14G}
showed that constraining the value of the cosmological constant,
$\Lambda$, was feasible using SNIa data, if the surveys targeted a
wide range of redshifts, thereby allowing to break the degeneracy in
cosmological parameters.

\begin{figure}[h]
\epsfxsize=\textwidth
\centerline{\epsfbox{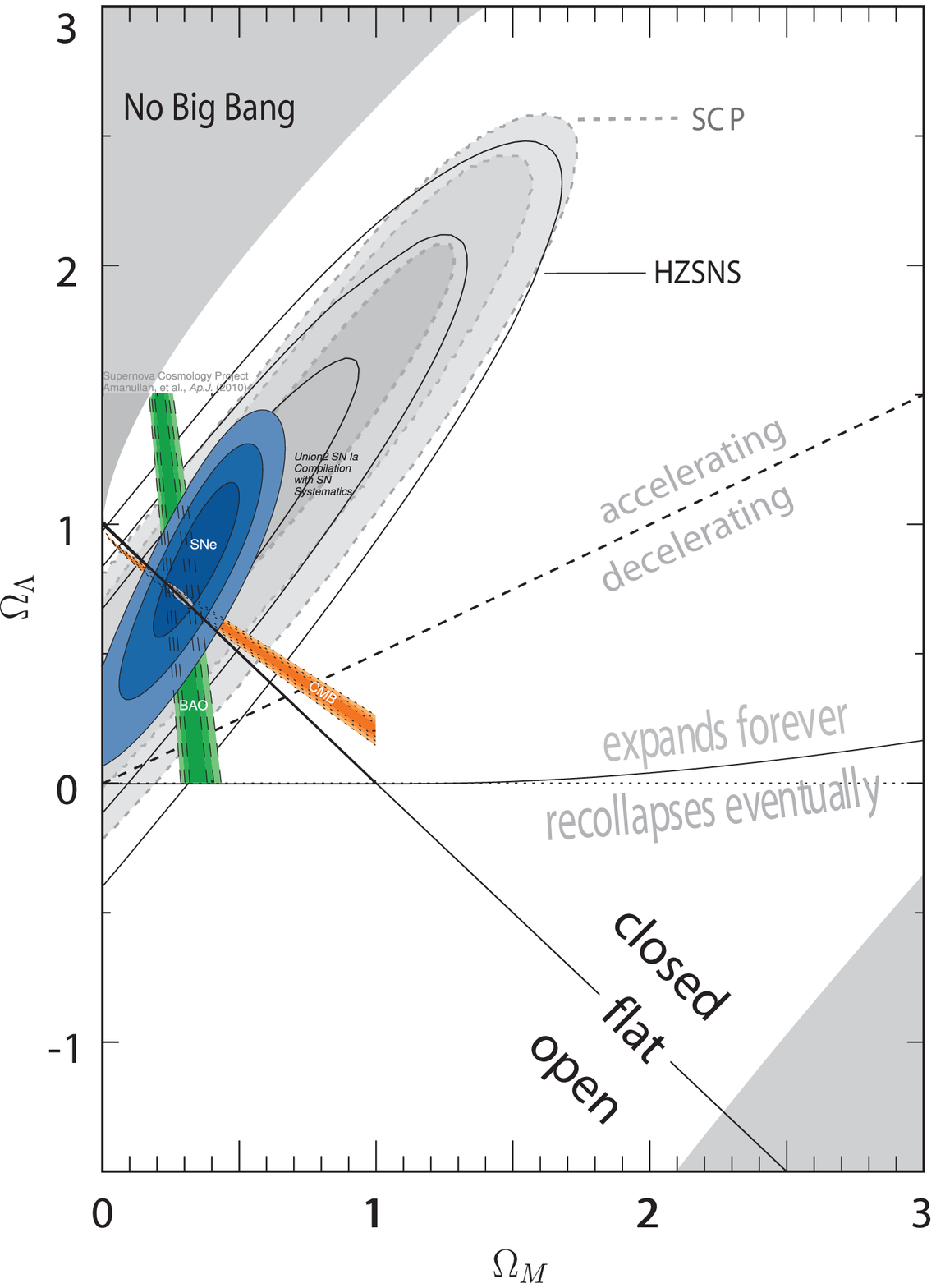}   }
\caption{
Confidence level regions in the 
$(\Omega_M,\Omega_\Lambda)$ parameter plane. The larger contours show
the original results from the High-Z Supernova Survey (HZSNS;
\cite{1998AJ....116.1009R}) and the Supernova Cosmology
Project (SCP; \cite{1999ApJ...517..565P}). The blue areas
correspond to the 68.3, 95.4 and 99.7\% confidence limits of the 
Union2 compilation of SNe~Ia \cite{2010ApJ...716..712A}.
Baryonic acoustic oscillations (BAO; green)
and cosmic microwave background (CMB; orange) constraints from \cite{2010MNRAS.401.2148P} and 
\cite{2010arXiv1001.4538K} are also shown, along with the combined 
contours from the three probes. Figure adaptation by R.~Amanullah.
The dramatic reduction in the confidence area without a change of the
most likely parameters is noteworthy and validates the early results.}
\label{fig:acceleration}
\end{figure}

Figure \ref{fig:acceleration}
shows the $\Omega_M-\Omega_\Lambda$ plane, with the two original
results of the SCP \cite{1999ApJ...517..565P} and HZ SN Team
\cite{1998AJ....116.1009R}, along with the recent Union2 compilation
based on 557 SNIa \cite{2010ApJ...716..712A}. The data-set includes
192 SNe with $z \le 0.15$  from
\cite{1996AJ....112.2398H,1999AJ....117..707R,2006AJ....131..527J,2008ApJ...686..749K,
2009ApJ...700..331H}. The intermediate and high-$z$ includes samples
from
\cite{1998AJ....116.1009R,1999ApJ...517..565P,2003ApJ...594....1T,
2003ApJ...598..102K,2005AJ....130.2453K,2004ApJ...602..571B,
2006A&A...447...31A,2008A&A...486..375A,2007ApJ...659...98R,2007ApJ...666..674M,
2008AJ....136.2306H}, adding up to 106 objects with $0.15 < z \le
0.55$, 64 with $0.55 < z \le 0.75$, 41  with $0.75 < z \le 0.95$ and
28 SNIa with $z>0.95$. The SN contours are derived from the Hubble
diagram data shown in Figure \ref{fig:Union2}.

\section{Cross-cutting techniques}
\label{sec:other}
\begin{figure}[h]
\epsfxsize=\textwidth
\centerline{\epsfbox[1 170 640 660]{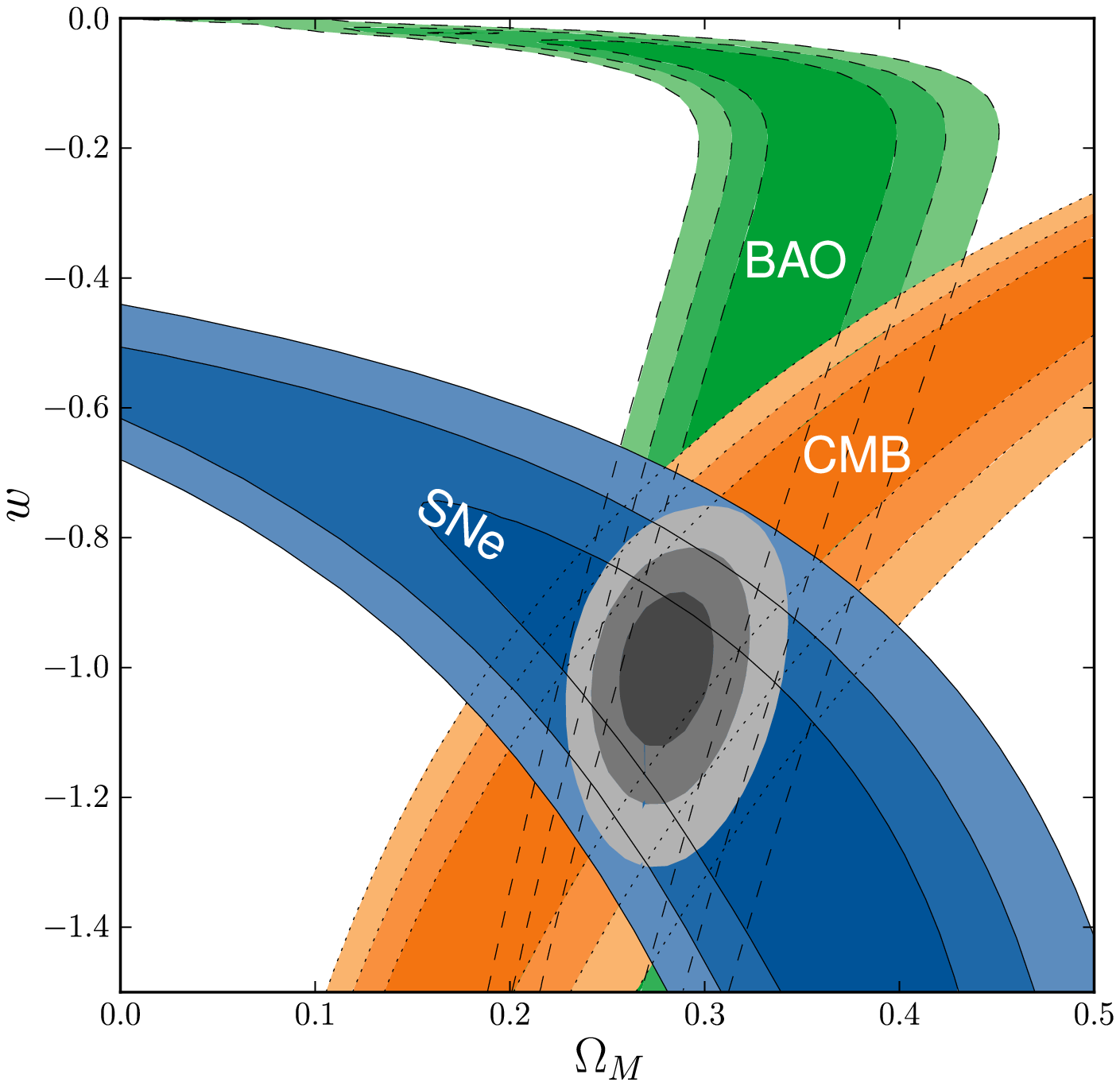}  }
\caption{
The current best constraints for a constant $w$ are shown here. A
cosmological constant ($w$=0) still is valid solution. The
complementarity of the different methods is apparent. The supernova
probability distribution is nearly perpendicular to the ones from the
baryonic acoustic oscillations or the microwave background. Data taken
from \cite{2010ApJ...716..712A}.
}
\label{fig:w}
\end{figure}

The power of the SNIa data for constraining cosmological parameters is
vastly enhanced when the likelihood functions are combined with what
is found with other probes, most notably the cosmic microwave
background (CMB) anisotropies and baryon acoustic oscillations (BAO),
as shown in Figure \ref{fig:acceleration} and discussed in
\cite{2008ARA&A..46..385F}. The combination of the different techniques 
becomes dramatically clear in the determination of $w$. For this
measurement all methods show large degeneracies, but their combination
provides a powerful tool to constrain a constant $w$, as shown 
in Fig.~\ref{fig:w}.

Similarly to the use of SNIa as ``standard candles'' for distance
estimates, the CMB and BAO techniques rely on ``standard rulers''.
Matter-radiation oscillations with wavelengths given by the product of
the sound speed and the elapsed time since the Big Bang leave a
characteristic bump in the power spectrum of the CMB temperature
fluctuations originating at $z\sim 1100$ and corresponding to an
angular scale of about 1$^{\circ}$ on the sky today. Measurement of
this angular scale in WMAP 7-year data \cite{2010arXiv1001.4538K}
leads to the precise constraints cosmological parameters, as shown in
Figs. ~\ref{fig:acceleration} and \ref{fig:w}. As the universe expands and photons
decouple from baryons, the matter anisotropies grow into the
structures observed today, down to the scale of galaxies and clusters
of galaxies. By studying a sample of Luminous Red Galaxies in the SDSS
survey, Eisenstein \etal~\cite{2005ApJ...633..560E} could
measure the $\sim 150$~Mpc length of the ``ruler'' at $z\sim 0.35$
leading to significant parameter constraints, mainly on $\Omega_M$.

Whereas the SNIa Hubble diagram is anchored at $z\approx0$, the CMB
and BAO originate in the early universe. As a result, the
``orientation'' of the confidence regions differ between the
techniques, thus rendering the crossing of the allowed regions
particularly useful to derive accurate cosmological parameters.

\section{The era of precision cosmology}
\label{sec:precis}
The pioneering efforts by the SCP (led by Saul Perlmutter) and the 
High-Z team (led by Brian Schmidt) in the mid and late 90's were
followed by new SN programs which developed the field in several
directions. 

Two major programs were established to
collect large samples of high quality data in the $z=[0.3-1]$ redshift
range: the ESSENCE program at the CTIO 4-m in Chile
\cite{2007ApJ...666..694W} and SNLS at the 3.6-m CFHT in Hawaii
\cite{2006A&A...447...31A}. 
These programs were
backed up with significant spectroscopic identification efforts from
$8-10$~meter class telescopes: VLT, Keck, Gemini and Magellan. The
3-year SNLS sample recently reported by \cite{2010A&A...523A...7G}
includes over 250 SN Ia, i.e., about 180 additional supernovae to what
is already included in the Union2 sample. SNLS and ESSENCE have
additional $\sim$300~SNe~Ia unpublished so far. 

One of the observational breakthroughs of SNLS was introducing the
``rolling search'' technique. One-square-degree fields were revisited
every 4 to 5 nights in multiple bands, thus new objects were found while
light curves were built of previously discovered supernovae. The
excellent time sampling with several filters and with a single
telescope and CCD system dramatically improved the quality of the SN
light curves.

The redshift ``gap'' around $z\sim0.2$ is filled by the SDSS-II
supernova survey. The first data \cite{2009ApJS..185...32K} includes
103 SNe~Ia with redshifts $0.04 < z < 0.42$. Overall, it is expected
that the SDSS-II supernova survey has about 400 supernovae for the
analysis.

Finally, the Hubble diagram is anchored with supernovae from the
Calan-Tololo \cite{1996AJ....112.2398H} and CfA surveys
\cite{1999AJ....117..707R,2006AJ....131..527J, 2009ApJ...700..331H}
(see also the list of nearby projects given in Sec.~\ref{sec:class}).

The HST program led by Adam Riess
\cite{2004ApJ...607..665R,2007ApJ...659...98R} dramatically enlarged
the observed redshift range of SNe~Ia. Thanks to the near-IR
capabilities from space, the $z>1$ domain was opened, allowing the
study of SNe in the decelerated phase of the expansion, i.e. exploring
the transition between acceleration and deceleration marked by the
turn-around of the differential Hubble diagram in the bottom panel of
Figure \ref{fig:Union2}.   

Progress at $z>0.9$ is slower, mainly because it hinges on major 
access to near-IR observations with HST. The SN detection 
efficiency can be improved by targeting massive galaxy clusters, as 
demonstrated by \cite{2009AJ....138.1271D}. Currently, a massive
project to find very distant ($z>1.5$) supernovae has started making
use of two multi-cycle HST projects, which image several areas and
galaxy clusters, over several years. Yet, it is unlikely that the
number of $z>1$ SNe~Ia that can be used in the Hubble diagram will
exceed $\sim$ 100 with current instruments.




The cosmological results are summarized in Table
\ref{tab:cosmo}. 


\begin{table}
\begin{tabular}{|l|l|l|l|l|} \hline
Ref & N$_{\rm SN}$ & $\Omega_{M}$ (flat) & $w$ (constant, flat) & LC fitter \\ \hline \hline
Astier \etal \, (2006) & $115$ & 
$0.263^{+0.042}_{-0.042}\ ^{+0.032}_{-0.032}$ &$-1.023^{+0.090}_{-0.090} \ ^{+0.054}_{-0.054}$ & SALT \\ \hline

Wood-Vasey \etal \, (2007) & $162$ & $0.267^{+0.028}_{-0.018}$ &
$-1.069^{+0.091}_{-0.093} \ ^{+0.13}_{-0.13}$ & MLCS2k2 \\ 

 \  & $178$ & $0.288^{+0.029}_{-0.019}$ &
$-0.958^{+0.088}_{-0.090} \ ^{+0.13}_{-0.13}$ & SALT2 \\ \hline

Kessler \etal \, (2009) & $288$ & $0.307^{+0.019}_{-0.019} \
^{+0.023}_{-0.023}$ & $-0.76^{+0.07}_{-0.07} \ ^{+0.11}_{-0.11}$ &
MLCS2k2 \\ 

 \  & $288$ & $0.265^{+0.016}_{-0.016} \
^{+0.025}_{-0.025}$ & $-0.96^{+0.06}_{-0.06} \ ^{+0.13}_{-0.13}$ &
SALT2 \\ \hline

Amanullah \etal \, (2010) & $557$ & $0.279^{+0.017}_{-0.016}$ &
$-0.997^{+0.050}_{-0.054} \ ^{+0.077}_{-0.082}$ & SALT2 \\ \hline

\end{tabular}
\caption[]{Cosmological parameters from the first year papers of
SNLS, ESSENCE and SDSS and the Union2 sample, Refs.~\cite{2006A&A...447...31A,
2007ApJ...666..694W,2009ApJS..185...32K,2010ApJ...716..712A}. The fits
include lower redshift data as well as BAO and CMB information. A flat
$w-$CDM model with constant $w$ is assumed in all analysis (see text
for more detail). }
\label{tab:cosmo}
\end{table}

The results are compatible and the strongest deviations come from
the use of the light curve fitters (and their assumptions on
reddening; see Sec.~\ref{sec:lcfit}). It should also be pointed out
that the samples are not completely independent. 
The nearby supernova sample for almost all results presented in
Tab.~\ref{tab:cosmo} are identical. The Union2 data set is the
collection of all published SN~Ia data suitable for cosmology.


\section{Systematic uncertainties}
\label{sec:systematics}
\subsection{Calibration}
Precise distance estimates with Type Ia supernovae require very
accurate relative photometry ($\Delta w \approx 2\Delta m$). While
future surveys are likely to measure most (all) their SNe with a
single, well calibrated system, current cosmological bounds are
derived from a compilation of surveys. There is a worry for
calibration offsets between the data sets, potentially resulting in a
bias in the fitted cosmology. The SNLS team has been extremely
active in refining the calibration of the CFHT MegaCam system and have 
reached sub-percent precision in their photometric
calibration \cite{2009A&A...506..999R}. Unfortunately, significant calibration 
uncertainties remain for the  
near-IR HST system, including NICMOS nonlinearities. This amounts to 
a sizable contribution to the budget of systematic errors 
for the $z\gsim 1$ SNe~Ia measured up to now \cite{2011ApJS..192....1C}.
Clearly, future SN~Ia programs must integrate a robust calibration program
into the survey design. 

\subsection{The UV-flux}
\label{sec:UV} 

Since high-$z$ surveys are mostly confined to optical bands, the
measured (redshifted) fluxes originate at ultraviolet and blue
wavelengths. For restframe wavelengths shorter than what corresponds
to the $B$-band, several sources of concern have been raised. The
atmospheric cut-off and steeply falling instrument
sensitivity has lead to significant difficulties in providing a good
estimate of the intrinsic brightness in the $U$-band.
About 5\% of the SNIa flux is emitted at wavelengths shorter than the
$U$-band, not at all accessible from the ground for $z=0$. The wavelength
region below $3000\,${\AA} remains one of the least explored, in spite
of the fact that it is rich in key information, e.g., about the
underlying explosion physics. 
Furthermore, since the dimming by dust grains increases rapidly at
shorter wavelengths, the uncertainty in host galaxy dust extinction
properties and intrinsic color variations are major limitations for
precise distance measurements. 


Photometric studies with the
{\em Swift} satellite  \cite{2010ApJ...721.1608B, 2010ApJ...721.1627M}
and STIS spectroscopy with HST \cite{2010arXiv1010.2211C} have been
used to test model predictions suggesting an increased dispersion in 
the UV, because of its larger sensitivity to progenitor composition
\cite{2000ApJ...530..966L,2000ApJ...528..590H,2008MNRAS.391.1605S}. These
studies, as well as similar ones at $z\sim 0.5$
\cite{2008ApJ...674...51E} do indeed find a scatter about twice as
large in the $UV$ compared to the optical.

It remains unclear what these effects mean for the use of SNIa as distance
indicators.

\subsection{Reddening and absorption}
\label{sec:reddening}
Accurate corrections for absorption of the SN light along the path from
the source to the telescope are crucial for distance determinations.
Interactions of light and dust grains may happen in the circumstellar
material around the exploding star, in the interstellar medium of the
host galaxy, in the intergalactic medium and in the Milky Way.
Absorption in the Earth's atmosphere is corrected through the
photometric calibration. The
distribution and properties of dust grains in the Milky Way is known
well enough to allow for correction accurate at the sub-percent level.
The other terms remain more problematic. The dust size-distribution
is captured by the Galactic extinction law and is often described by
the total-to-selective extinction ratio, $R_V$. 
For the Milky Way, an average value $R_V=3.1$ has been found
\cite{1989ApJ...345..245C}, with relatively small scatter around this
value across the galaxy. The existence of significant amounts of
intergalactic dust, $\Omega_{dust}\sim 10^{-6}-10^{-5}$, has been
inferred from estimates of the stellar density and metallicity as a
function of redshift. After the evidence for cosmic acceleration was
presented in 1998, several authors studied the possibility that the
faintness of SNe~Ia at high-$z$ could instead be caused by ``gray''
interstellar dust in the intergalactic medium
\cite{1999ApJ...512L..19A,1999ApJ...525..583A, 2002A&A...384....1G}.
An upper limit on the restframe $B$-band attenuation $A_B(z=1) < 0.1$
on impact of dimming by dust in the intergalactic medium was derived
by studies of quasar colors
\cite{2003JCAP...09..009M,2005JCAP...02..005O}.

More recently, Menard \etal \cite{2010MNRAS.405.1025M} performed a
cross-correlation study between colors of nearby and distant quasars
and reported a statistical detection of dust reddening up to large
scales around galaxies. From this result, they inferred the opacity of
the universe.  Their model-dependent estimate gives $A_B(z=0.3)
\approx  0.02$ mag.  This attenuation is partially compensated by the
color-brightness correction. However, since the attenuation happens
along the line of sight rather than at the redshift of the SN, the
current light curve fitters cannot account for this effect in its
totality as they correct for discrete redshifts of the host galaxy.

Considerably larger extinction by dust may occur in the interstellar
medium of the host galaxy, especially for SN explosions close to the
galaxy center. Multi-band photometry of SNe~Ia has been used to
constrain the wavelength dependence of the absorption.  There is by now a
large number of studies reporting measurements of $R_V \sim 2$ based
on SN~Ia colors, (e.g.
\cite{2006AJ....131.1639K,2006MNRAS.369.1880E,2007A&A...466...11G,
2007AJ....133...58K,2008A&A...487...19N,2008MNRAS.384..107E}).
Moreover, low-values of $R_V$ do also provide a smaller scatter in the
Hubble diagram and thus a better fit in global parameter fitting of
SNe~Ia across a wide redshift range. This is somewhat puzzling and
indicates that reddening by dust is not fully understood. 

Multiple scattering of photons has been suggested as possible
explanation to the ``unexpected'' reddening law. This could be the
case in the presence of dust in the circumstellar environment of the
SN~Ia progenitor, as suggested by  \cite{2008ApJ...686L.103G}, who also
provided an effective reddening law that gives good fits to the best
measured/most reddened SNe~Ia in the CSP low-$z$ sample
\cite{2010AJ....139..120F}. Spectroscopic evidence for circumstellar
material around highly reddened SNe Ia has been found in e.g.
\cite{2007Sci...317..924P}.

At this point, the lack of understanding of the relative strengths and
specific properties of the various effects contributing to the
color-brightness relation of SNe~Ia constitutes one of the biggest
concerns, especially since a redshift dependence cannot be excluded.

\subsection{Light curve fitters} 
\label{sec:lcfit}
\begin{figure}[h]
\epsfxsize=\textwidth
\centerline{\epsfbox[1 1 470 330]{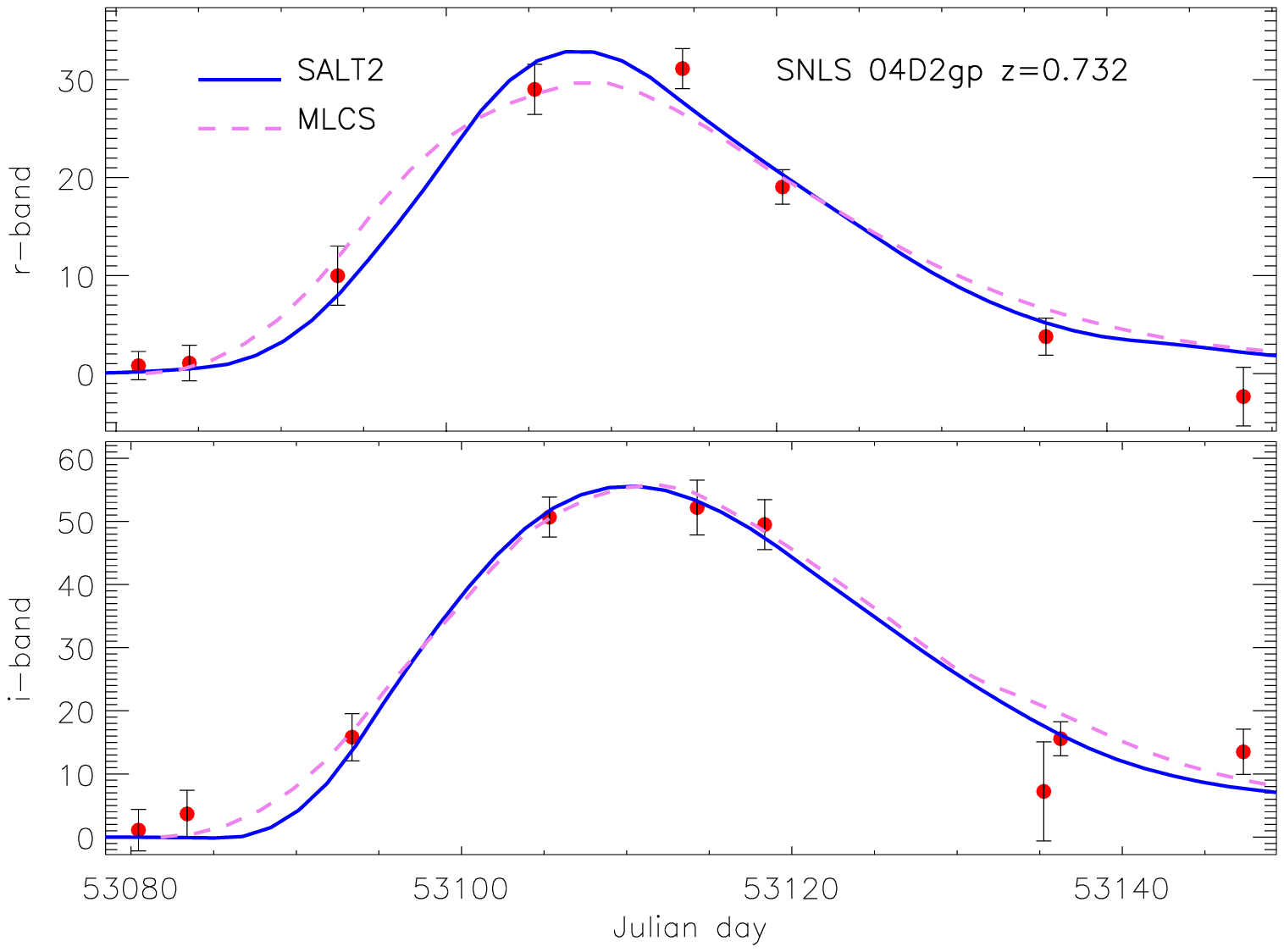} }
\caption{
The differences incurred from using different light curve fitters is
demonstrated here. The same $r-$ and $i$-filter observations 
(fluxes, not magnitudes) of the
SNLS supernova 04D2gp (z=0.732) are
approximated with the SALT2 and the MLCS2k2 fitters, two of the most
used fitters at the moment. Small deviations in the $i-$filter fitting
can be seen and are due to the different underlying templates used in
the fitters and differences in how the colors are determined
simultaneously. 
}
\label{fig:lcfit}
\end{figure}

Establishing the peak brightness of SNe~Ia has to date been
essential essential for their use in
cosmology. A major progress over the past decade has been the much
more systematic sampling of supernova observations. While the light
curves of early supernovae had to be established from rather sparsely
sampled data, the current and future surveys provide a much denser
temporal coverage and the determination of the peak brightness is
highly improved. 
However, light curve shape and reddening have to be established
through fitting. 
Since color and shape are measured with respect to some reference set
and may also include some additional priors, several light curve
fitters have been proposed over the years. Examples are the fitter
developed by the SCP, SNMinuit, used the original light curve shapes in
\cite{1988PhDT.......171L} based on low-$z$ data, and improved in
\cite{2001ApJ...558..359G}. A ``stretch'' factor, $s$, was introduced
to characterize the light curve shape perturbations with a simple
scaling of the time axis within the, $t \cdot s$, and a
corresponding brightness correction $\Delta M= \alpha\cdot (1-s)$,
i.e. broader light curves $s>1$ corresponding to brighter intrinsic
brightness, $\Delta M<0$ (cf. Fig.~\ref{fig:SNcalibration}). While $s$ is fitted with points both in the
rising and falling parts of the light curves, other shape estimates,
like $\Delta m_{15}$ \cite{1999AJ....118.1766P}, use only a limited
region after maximum. The two major LC fitters used today are
SALT2\cite{2007A&A...466...11G} and MLCS2k2
\cite{2007ApJ...659..122J}, the second generations of SALT
\cite{2005A&A...443..781G} and MLCS \cite{1996ApJ...473...88R}. The
major differences between the two reside in the reference spectral
templates in the blue and $UV$ regions  and in the interpretation of
reddening.
While the MLCS approach is to obtain the reference samples from the
low-$z$ SNe, SALT combines data from both high and low redshift. In
consequence, the MLCS ``training'' may be biased by the difficulties
to obtain high quality data in the $UV$, as explained in section
\ref{sec:UV}, whereas SALT may lack the ability to detect intrinsic
changes with redshift and makes use of faint data. Another key
difference involves the treatment of color-brightness relation. MLCS
assumes that SN~Ia reddening is caused by dust, thus negative color
corrections are not used. In contrast, SALT combines all possible
contributions into a single function relating the intensity as a
function of wavelength, and both positive as well as negative
corrections are allowed, i.e. a negative color excess implies
increased brightness and reddening corresponds to dimming of light.
An example of MLCS2k2 and SALT2 fits of the SNLS supernova 04D2gp
($z=0.732$) is shown in Figure \ref{fig:lcfit}. Although both fits
give good $\chi^2$, SALT2 finds the SNe to be moderately blue, while
MLCS2k2 puts it at small (but positive) extinction. Thus, the derived
distance modulus for this particular SN differs by 0.1 mag
\cite{2009ApJS..185...32K}. More recently, two other fitters have been
developed SiFTO \cite{2008ApJ...681..482C}, so far only used by the
SNLS collaboration, and SNooPy \cite{2011AJ....141...19B} by CSP that,
unlike other fitters, can also be used to build near-IR light curves. 

\subsection{Selection bias}
\begin{figure}[h]
\epsfxsize=\textwidth
\centerline{\epsfbox[1 1 555 280]{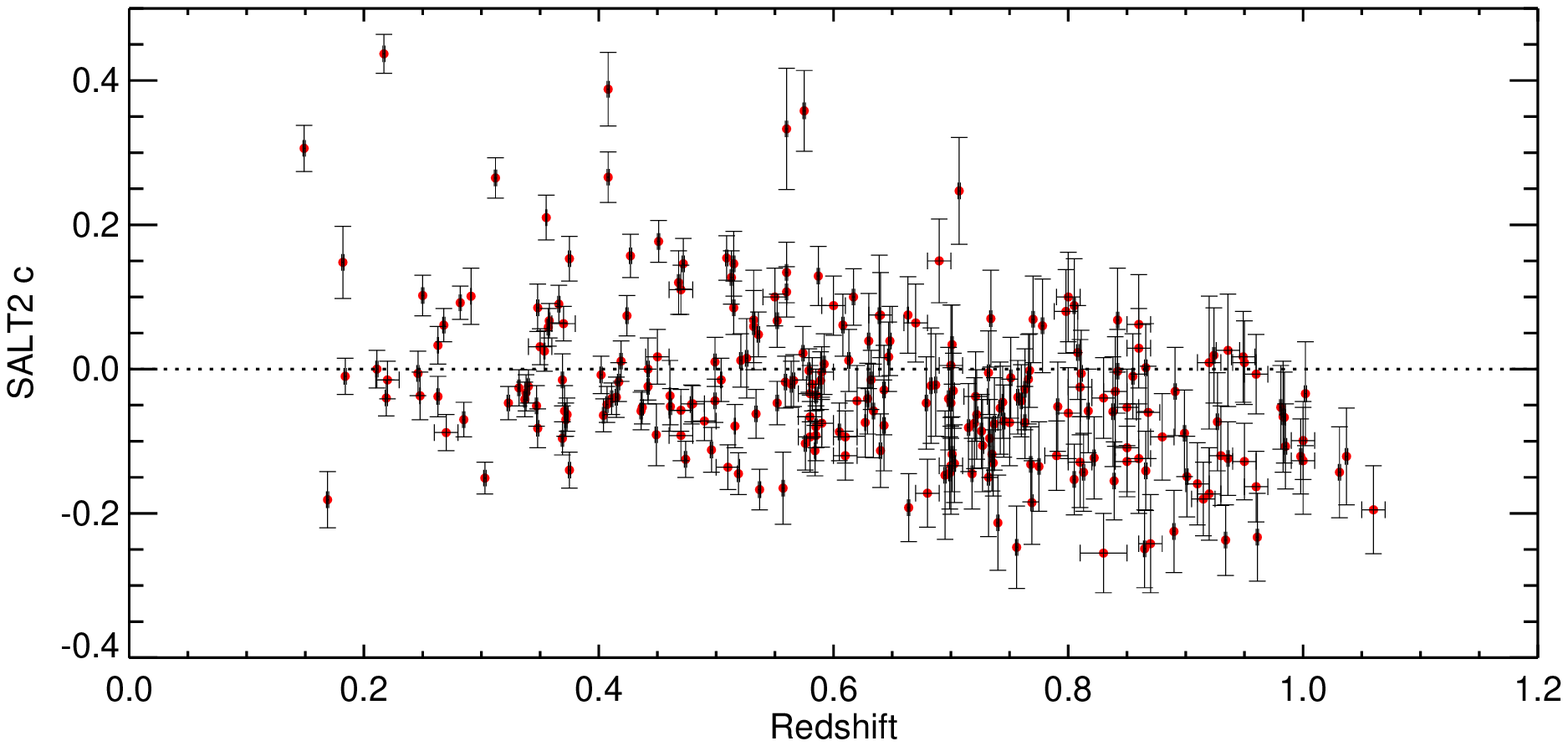}}
\caption{Fitted SALT2 color, $c$, vs redshift for the 3-year
SNLS data-set \cite{2010A&A...523A...7G}. There is strong evidence
for a selection effect,  with bluer supernovae being found at larger
redshifts. Since bluer supernovae are intrinsically more luminous this
effect can be explained as a selection driven by the magnitude limit of
the spectroscopic follow-up.}
\label{fig:spectrobias}
\end{figure}

Varying selection efficiency is an inherent difficulty in a technique
involving comparisons of brightness of objects at different redshifts. 
In particular, cosmological estimates using multiple
surveys need to model (most often through Monte Carlo simulations) the
fraction of the SN~Ia luminosity function sampled by the detection
requirements of each instrument. At high-$z$, the ``bottle-neck'' is
the spectroscopic identification, which is only practically obtainable
for $i<25$ mag in 8-10 meter class telescopes. This naturally leads to
a redshift dependent selection bias in magnitude and color, e.g., 
for the 3-year SNLS sample \cite{2010A&A...523A...7G}, as shown
in Figure~\ref{fig:spectrobias}. Perrett \etal 
\cite{2010AJ....140..518P} examine these selection
effects in detail, and the related corrections. A problem associated
with these corrections is our lack of precise understanding of 
potential drifts in either reddening properties (Sec.~\ref{sec:reddening})
or intrinsic SN~Ia luminosity, to be discussed further in 
Sec.~\ref{sec:evolution}. An ambitious space mission equipped with 
a multi-band imager and a
sensitive spectrograph could potentially resolve this difficulty
(see Sec.~\ref{sec:stage4}).

\subsection{Gravitational lensing}
\label{sec:lensing}

\begin{figure}[h]
\epsfxsize=\textwidth
\centerline{
\epsfbox[90 90 520 520]{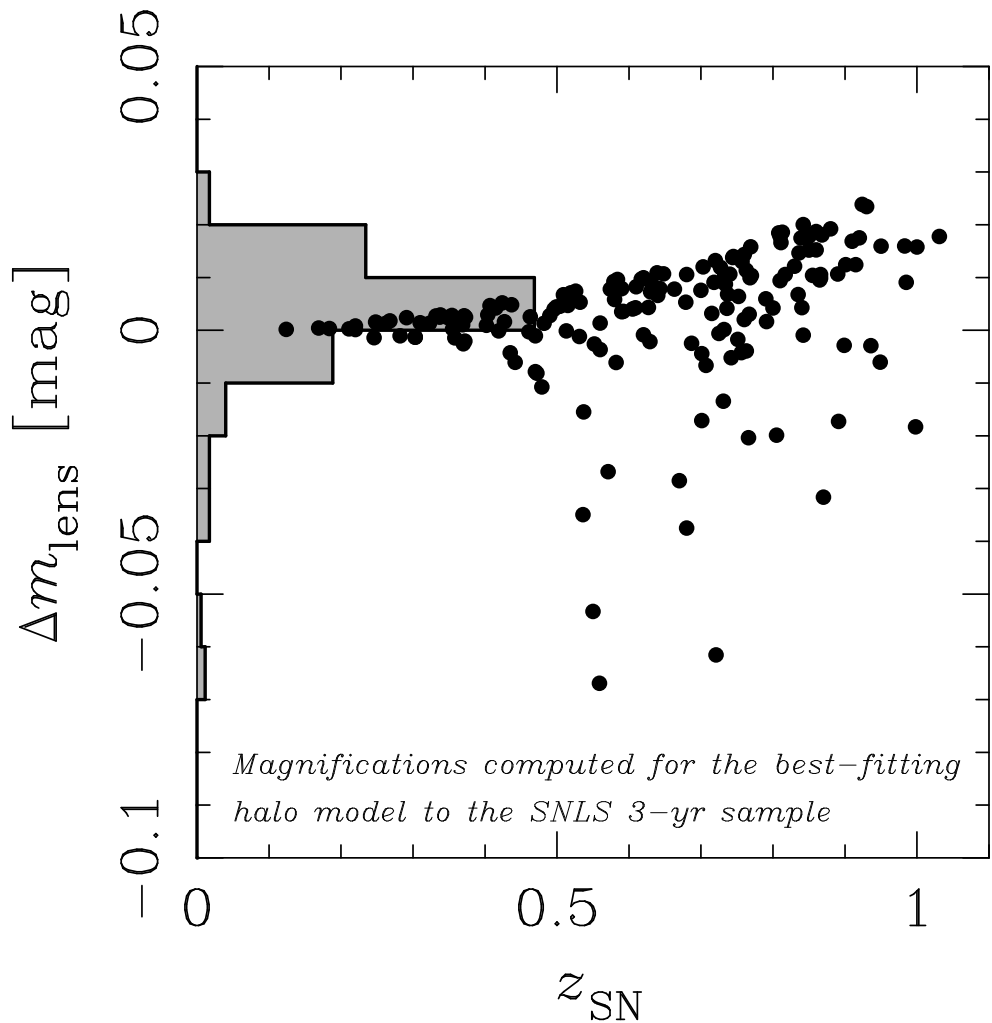}
}
\caption{Magnification of SNe~Ia in the 3-year SNLS data 
vs.~redshift. 
Filled circles represent the estimated lensing magnification, 
$\Delta m_{lens}$, 
to individual Type Ia SNe, based on the best fit halo model
from \cite{2010MNRAS.405..535J}. The effect grows with redshift, but the 
mean value is close to zero, as also indicated by the histogram on the
side. Figure courtesy of Jakob J\"onsson.}
\label{fig:lensing}
\end{figure}

In Section \ref{sec:Friedmann} we derived the expression for the
luminosity distance used to measure cosmological parameters assuming
that the universe is homogeneous and isotropic. Although there is
ample empirical evidence that this is a good approximation at the
largest scales, e.g., from the cosmic microwave background, it is
clear that inhomogeneities exist at smaller scales. Thus, foreground
galaxy halos and clusters of galaxies act as gravitational lenses and
introduce a ``scatter'' in the observed SN~Ia flux that increases with
redshift
\cite{2000A&A...358...13B,2005ApJ...631..678H,2001ApJ...561..106M}.
Since the number of photons is conserved under lensing (some objects
get magnified and others de-magnified), the luminosity distance
averaged over a large number of SNe~Ia is expected to converge towards
Eq.~(\ref{eq:dl1}).  However, for small number statistics or if the
sample is magnitude limited, gravitational lensing could be a source
of selection bias. This was first investigated in
\cite{2006ApJ...639..991J} for the HST GOODS sample
\cite{2004ApJ...607..665R}, where the redshifts of the foreground
galaxies are well measured and their masses can be estimated through
their luminosities and the effect of lensing of individual SN can be
evaluated. More recently, J\"onsson \etal \cite{2010MNRAS.405..535J}
have studied the weak lensing perturbations in the SNLS 3-year data
(Figure~\ref{fig:lensing}). The distribution of lensing magnifications
of SNe~Ia as a function of redshift matches the expectations and no
significant bias has been detected in the current high-$z$ sample. 

\subsection{Peculiar velocities}
Local matter inhomogeneities are also important for precision
cosmology with SNe~Ia. Local
density perturbations induce peculiar velocities in galaxies of about
300 km s$^{-1}$. In addition, the exploding star could have a
comparable internal motion within the host galaxy. Since these
velocity components cannot be disentangled from the cosmological
redshift, an extra source of uncertainty
needs to be considered, 
$\sigma_{\mu} 
\approx 
{v \over c} \cdot{5 \over \ln(10)}\left[ {1+z}\over {z(1+z/2)} \right],
$ which exceeds the error from intrinsic brightness $\sigma_{\mu} \approx
0.1$ mag for $z<0.015$. The concern, however, is not the addition of
an added ``random'' error, but rather the possibility of an effect
that does not average out with statistics.
In particular, our own motion and the correlated
peculiar velocities of the SN~Ia host galaxies at low-$z$ could have a
non-negligible effect on the derived cosmological parameters. Several
authors have studied these effects, concluding that inclusion of
SNe~Ia down to $z=0.02$ may result in a systematic uncertainty of
about $\Delta w \lsim 0.01$ \cite{2010arXiv1012.2912D}.

\subsection{Brightness evolution} 
\label{sec:evolution}
The brightness distribution of SNe~Ia has been shown to depend on the
host galaxy environment. E.g. Hamuy \etal
\cite{1996AJ....112.2391H} noticed that the
average B and V-band light curve width differs between galaxy types. It
is now an established fact that elliptical galaxies host a larger
fraction of narrower, low ``stretch'', supernovae. 
Furthermore, recent studies of nearby supernovae
\cite{2010ApJ...715..743K}; SNLS supernovae \cite{2010MNRAS.406..782S}
and SDSS SNe \cite{2010ApJ...722..566L} suggest possible evidence for 
varying supernova properties with host
galaxy not entirely corrected by the standard width-color-brightness
relations. In particular, it has been argued that splitting the
sample according to host-galaxy mass and allowing the peak absolute
magnitudes of the two samples to differ improves the residuals
in a statistically significant manner. The evidence
suggests that large stellar mass galaxies host the brightest SNe Ia
($\Delta M \sim 0.075$ mag), after
light curve shape corrections. 
Given the time evolution of metallicity and stellar mass of galaxies, 
the possibility of a drift in the brightness of the ``standard candle''
with redshift remains a source of concern for precision cosmology.

\subsection{The ``averaging'' uncertainty}
After the discovery of  dark energy several authors
\cite{2000PhRvD..62d3525B, 2007astro.ph..2416C,2009PhRvD..80l3512W,
2010arXiv1012.0784R,2010GReGr..42.1399K} 
raised doubts about the theoretical foundations leading to
Eq.~(\ref{eq:dl}). In particular, it has been claimed that the average
expansion in a locally inhomogeneous universe behaves differently than
expected in Friedmann-Lemaitre models based on the large scale {\em
average} energy density. The reason being the non-linearity of the
Einstein equations: spatial averaging and solving the Einstein
equations do not commute \cite{1984grg..conf..215E}. This is sometimes referred to
as the ``back-reaction'' problem. Although the issue is
generally accepted, arguments have been put forward for why the effect is
too small to explain ``cosmic acceleration'' without dark energy.  In
a matter-dominated universe the local dynamics should be (almost)
Newtonian, as long as potential perturbations and peculiar velocities
are non-relativistic. Since Newtonian gravity is linear, averaging and
evolution do commute in the Newtonian limit and should commute to a
good approximation in general relativity \cite{2006CQGra..23..235I,2005PhRvD..71j3521F}.  
Any relativistic
corrections should be much to small to mimic effective acceleration.  
Since it is notoriously difficult to estimate
the size of the effects from non-linear back-reaction in general relativity,
it remains unclear at this point whether this presents a limitation for
precision tests of $w$.

\section{Recent developments}
\label{sec:recent}
\subsection{Expanding the wavelength window}
The use of Type Ia supernovae helped establish a major milestone
in precision cosmology, especially the discovery of dark energy. However,
further progress in the field is limited by our ability to resolve the
systematic uncertainties outlined in Section \ref{sec:systematics}. Some
outstanding issues involve understanding the progenitor system
leading to Type Ia supernovae and the properties of dust extinction
affecting SN~Ia. A natural path that has been followed for this purpose is
to expand the studied wavelength window, especially for nearby SNe.
Furthermore, theoretical arguments have been put forward suggesting that
the dispersion of intrinsic SNIa peak magnitudes should have a minimum
around 1.6$\mu$m \cite{2006ApJ...649..939K}.

\subsubsection{The near-IR}
The Earth's atmosphere severely limits the feasibility of ground 
based observations
in the near-IR. Vibrational bands of H$_{2}$0 and CO$_{2}$ block the
incoming radiation beyond 1 $\mu$m, leaving a few observational windows
in the near-IR: $J,H$ and $K$-bands, centered at $1.2, 1.6$ and $2.2$ 
$\mu$m, respectively. To make matters worse, significant thermal 
atmospheric emission (mostly from multiple narrow OH-lines) reduces 
the obtainable signal-to-noise ratio. This is particularly challenging
since the fractional flux in the near-IR is small for SNIa: 2, 1 and 0.5 \%
for $J,H,K$ \cite{2010arXiv1011.0441H}. In spite of these difficulties, 
significant efforts have been devoted recently to studies of
supernovae in the near-IR. A non-standard extinction law has been
measured with great accuracy for a number of highly reddened SNIa with
high S/N near-IR data \cite{2006MNRAS.369.1880E,2008MNRAS.384..107E,2007AJ....133...58K,2008ApJ...675..626W,2010AJ....139..120F}. Moreover, statistical
tests on $z<0.05$ SNIa in
 \cite{2008ApJ...689..377W,2009ApJ...704..629M,2010AJ....139..120F,2010arXiv1011.5910M} offer observational 
support for the claim in \cite{2006ApJ...649..939K}
of a somewhat narrower intrinsic scatter in the near-IR. However, this benefit may 
only be exploited at high-redshift with space based observations, e.g. 
the James Webb Space Telescope.
Because of the paucity of data in near-IR,
and the very narrow redshift range studied,  
it is yet unclear if problems with $K$-corrections may arise, i.e., varying
spectral features moving in and out of the filter pass-band as the redshift
increases. An interesting development taking place
in recent years, and more easily accessible than the restframe NIR, 
is building a Hubble diagram in restframe $I$-band, i.e., at about 8000 \AA \ \cite{2005A&A...437..789N,2009ApJ...700.1415N}.
In particular, Freedman \etal \cite{2009ApJ...704.1036F} use a sample 
of 56 SNIa; 21 nearby, and 35 in the redshift range $0.1 < z <0.7$, and find limits on $w$ which
are consistent with the results at shorter wavelengths. 

%


\subsection{Spectral indicators as secondary calibrators}
Thanks to the availability of large spectroscopic sets of SNIa, both
nearby and at cosmological distances, data mining efforts have been
conducted trying to identify spectral indicators capable of refining
the intrinsic dispersion of the ``standard candle'', e.g.,
\cite{2005AJ....130.2788H,2005ApJ...623.1011B,2006PASP..118..560B,2006AJ....131.1648B,2007AA...470..411G,2008ApJ...684...68F,2008A&A...477..717B,2008ApJ...674...51E,2009ApJ...693L..76S,2009A&A...500L..17B,2009ApJ...699L.139W,
2010MNRAS.tmp.1811W,2010arXiv1011.4517F,2010arXiv1011.6227N,2010arXiv1012.4430N,2011A&A...526A..81B}. Although some of these
findings could have a profound impact in our understanding of the physics
behind SNIa explosions and optimize their use in cosmology, the statistical
evidence and interpretation of the correlations needs further 
assessment.


\section{Supernova cosmology for the next decade}
\label{sec:outlook}
\label{sec:notlambda}
\label{sec:stage4}

The goal of the next supernova cosmology surveys must be to reach a
level of accuracy promising to meaningfully constrain a time-dependent
$w(t)$. From the descriptions of the systematics of the supernova
cosmology, it is clear that a strict control of the surveys is
required. This has led to the concentration on singular telescopes and
their calibration for the searches and the photometry. Finding
potential further correlations in the SN~Ia observations can further
sharpen the acuity of SNe~Ia for cosmological distances and will also
give hints on the supernova physics. Another improvement would be to
explore the small luminosity scatter of SNe~Ia in the near-infrared.
Building a well-sampled Hubble diagram at near-infrared wavelengths
would reduced the uncertainties on the reddening corrections as well. 

Several programs aiming at discovering thousands of SNe~Ia up to $z=1$
are in the process of starting up. The ``rolling search'' technique,
where CCD cameras with very large field of view are used to revisit
sky pointings with cadences of a few days, has been proved very
successful, e.g., by SNLS using the 1-square-degree at the 3.6m CFHT
in Hawaii. New SNe are found while multi-band light curves are built
for transients discovered in earlier epochs. With the imminent start
of operations of the Dark Energy Survey (DES) on the CTIO 4m (Chile)
using a new 3 square degree mosaic CCD camera, about 3000~SNe~Ia are
expected within 5 years. The redshift range will overlap with the
existing samples from SDSS, SNLS and ESSENCE. 


A fully developed Pan-STARRS with four 1.8m telescopes observing
simultaneously
will have the capability to scan the entire sky within just a week.
With the 3-square-degree CCD array on each unit and a pixel size of
0.3 arc-seconds, a large number of SNe~Ia can be found across a wide
redshift range. The first prototype unit, Pan-STARRS1, is already collecting
data and several supernova discoveries have been reported.

The 8-meter telescope foreseen in the LSST project, carrying a
9-square-degree camera and performing repeat all sky imaging in five
optical bands will propel the art of finding supernovae to a whole new
dimension. With an expected yearly yield of 250000 SNe~Ia, with light curves
sampled every five days, low- and high-$z$ supernovae will be found
with the same instrument and calibrated to about 1\% photometric
precision. 

The limitation of this project lies also in the big numbers: it is
impossible to get spectroscopic identification for more than a small
subset of supernovae. The redshift determination most likely will be
done through photometric redshifts of the host galaxies. Thus, the
success of the LSST SN~Ia program for cosmology depends on the accuracy
of the photometric identification and control of systematic
uncertainties related to e.g., a possible evolution of the standard
candle and brightness-color corrections, as discussed in
Sec.\ref{sec:systematics}.


First proposed in 1999, the SuperNova Acceleration Probe (SNAP) was
for a long time the leading concept for a dedicated satellite mission
aiming at the study of the properties of dark energy with SNIa.
Several competing designs have been proposed in the mean time,
focusing on one or several cosmological probes: Type Ia supernovae,
weak lensing measurements and baryon acoustic oscillations. At
present, two missions are being considered: the European-led EUCLID
satellite and a US-proposed WFIRST.
ESA's EUCLID is mainly focusing on baryonic acoustic oscillations and
weak lensing and the current specifications include only a broad
optical (``visible'') and three near-infrared filters, which is arguably
non-optimal for precision studies of SN~Ia over a wide range of
redshifts. The WFIRST mission concept includes SN~Ia (as well as
baryonic acoustic oscillations
and weak lensing) and a 1.5-meter class telescope has been proposed as a
reference. Although a space mission cannot compete with LSST in
statistics of SNe~Ia, it is likely to collect a much better studied
sample of SN~Ia, including near-IR observations and possibly
spectrophotometry, thereby having a better control of systematic
uncertainties, the ultimate limitation in studying the properties of
dark energy with SNIa.

\begin{figure}[h]
\epsfxsize=15cm
\centerline{\epsfbox{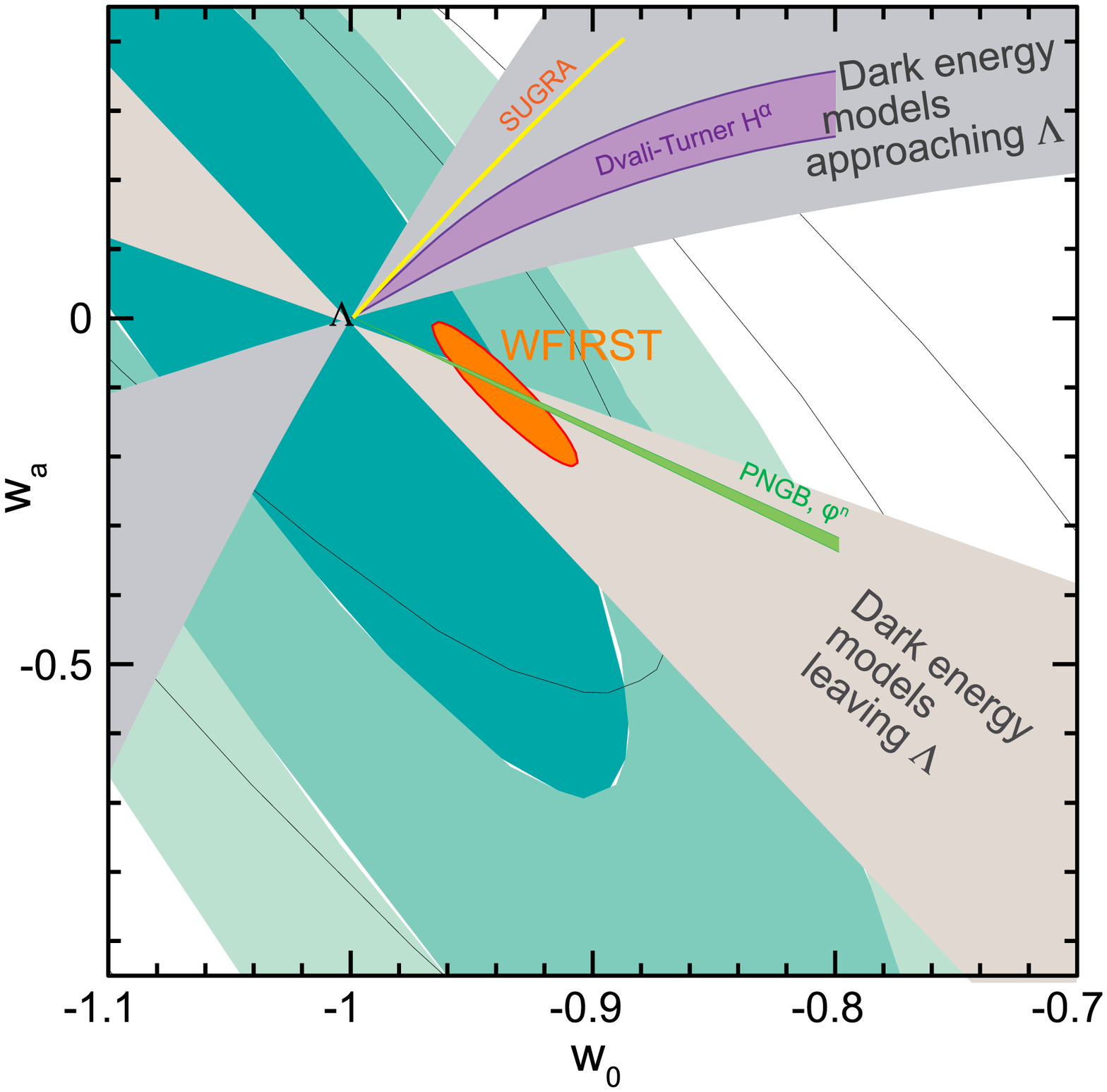} }
\caption{Projected 68\% confidence region in the $(w_0,w_a)$ plane
for a planned satelite mission (Eric Linder, private communication) 
compared with the Union2 results. While the existing data cannot
constrain a time-dependent $w$ future missions have the promise to
meaningful constrain this quantity. Figure adapted by Joel Johansson.}
\label{fig:wfirst}
\end{figure}

Figure \ref{fig:wfirst} shows the projected sensitivity of 
WFIRST in the $(w_0,w_a)$ parameter space
compared with the current constraints from the Union2 sample in
Amanullah \etal \,(2010) \cite{2010ApJ...716..712A}. The shaded
``petals'' in the figure show classes of ``quintessence'' models, i.e.,
where a scalar field dominates the current energy density of the 
universe \cite{2005PhRvL..95n1301C}.  One such case is the pseudo 
Nambu-Goldstone boson (PNGB) \cite{1995PhRvL..75.2077F}, e.g., an axion with 
a Hubble scale ($H_0^{-1}$) Compton wavelength, i.e., a mass 
$m \sim  10^{-33}$ eV. Predictions from 
String Theory inspired scalar field potentials 
in gravitational theories with extra dimensions (SUGRA) and the braneworld
model by Dvali and Turner \cite{2003astro.ph..1510D} are also shown for
reference. Clearly, {\em any} deviation from $\Lambda$ would be dramatic
breakthrough in fundamental physics. Observational cosmology, and distance
estimates with SNe~Ia in particular, will continue to contribute to
the full picture of high-energy and particle physics. 

\paragraph{Acknowledgements}
We would like to thank Rahman Amanullah, Andy Becker, Joel Johansson,
Jakob J\"onsson, Eric Linder, Rick Kessler and Max Strizinger for
providing material and help with some of the figures and
to Eric Linder for comments on the manuscript.  We are grateful
to the organizers and participants of the Aspen 2010 Type Ia supernova
workshop for helpful and stimulating discussions while parts of
this manuscript were written.  BL acknowledges support for this work by
the DFG through TRR33 "The Dark Universe".

\bibliography{GoobarLeibundgut}
\bibliographystyle{arnuke_revised}

\end{document}